\documentclass[sn-standardnature,iicol]{sn-jnl}

\usepackage{array}
\usepackage{graphicx}
\usepackage{multirow}
\usepackage{amsmath,amssymb,amsfonts}
\usepackage{amsthm}
\usepackage{mathrsfs}
\usepackage{xcolor}
\usepackage{textcomp}
\usepackage{manyfoot}
\usepackage{booktabs}
\usepackage{algorithm}
\usepackage{algorithmicx}
\usepackage{algpseudocode}
\usepackage{listings}
\usepackage{xspace}

\newcommand{\firefly}{\texttt{FIREFLy}\xspace}
\newcommand{\Juniper}{\texttt{Juniper}\xspace}
\newcommand{\POSEIDON}{\texttt{POSEIDON}\xspace}

\def\aap{{Astron.~Astrophys.}}
\def\apj{{Astrophys.~J.}}
\def\aj{{Astron.~J.}}
\def\apjl{{Astrophys.~J.~Lett.}}

\def\pasp{{Publ.~Astron.~Soc.~Pac.}}

\def\mnras{{Mon.~Not.~R.~Astron.~Soc.}}
\def\nat{{Nature}}
\def\icarus{{Icarus}}

\jyear{2026}%

\begin{document}

\title{\centering Aerosols and hydrocarbons in the atmosphere of a \\ white dwarf planet}

\author[1,2]{\fnm{Ryan} J. \sur{MacDonald}}
\author[3,4]{\fnm{Christopher} E. \sur{O'Connor}}
\author[3]{\fnm{Victoria} A. \sur{Boehm}}
\author[5]{\fnm{E.} M. \sur{May}}
\author[6]{\fnm{David} K. \sur{Sing}}
\author[3]{\fnm{Elijah} \sur{Mullens}}
\author[5]{\fnm{L.} C. \sur{Mayorga}}
\author[3,7]{\fnm{Trevor} O. \sur{Foote}}
\author[8]{\fnm{Simon} \sur{Blouin}}
\author[9,2]{\fnm{Logan} A. \sur{Pearce}}
\author[3]{\fnm{Nikole} K. \sur{Lewis}}
\author[10]{\fnm{Jeff} \sur{Valenti}}
\author[11]{\fnm{Natasha} E. \sur{Batalha}$^{\textrm{11}}$}
\author[3]{\fnm{Maura} \sur{Lally}}
\author[10]{\fnm{Joshua} D. \sur{Lothringer}}
\author[12]{\fnm{Mark} S. \sur{Marley}}
\author[13]{\fnm{Ishan} \sur{Mishra}}
\author[10]{\fnm{Susan} E. \sur{Mullally}}

\affil[1]{\orgname{University of St Andrews}, \orgaddress{ \city{St Andrews}, \country{UK}}}
\affil[2]{\orgname{University of Michigan}, \orgaddress{ \city{Ann Arbor}, \state{MI}, \country{USA}}}
\affil[3]{\orgname{Cornell University}, \orgaddress{ \city{Ithaca}, \state{NY}, \country{USA}}}
\affil[4]{\orgname{Northwestern University}, \orgaddress{ \city{Evanston}, \state{IL}, \country{USA}}}
\affil[5]{\orgname{Johns Hopkins Applied Physics Laboratory}, \orgaddress{ \city{Laurel}, \state{MD}, \country{USA}}}
\affil[6]{\orgname{Johns Hopkins University}, \orgaddress{ \city{Baltimore}, \state{MD}, \country{USA}}}
\affil[7]{\orgname{NASA Goddard Space Flight Center}, \orgaddress{ \city{Greenbelt}, \state{MD}, \country{USA}}}
\affil[8]{\orgname{University of Victoria}, \orgaddress{ \city{Victoria}, \state{BC}, \country{Canada}}}
\affil[9]{\orgdiv{Steward Observatory}, \orgname{University of Arizona}, \orgaddress{ \city{Tucson}, \state{AZ}, \country{USA}}}
\affil[10]{\orgname{Space Telescope Science Institute}, \orgaddress{ \city{Baltimore}, \state{MD}, \country{USA}}}
\affil[11]{\orgname{NASA Ames Research Center}, \orgaddress{ \city{Moffett Field}, \state{CA}, \country{USA}}}
\affil[12]{\orgdiv{Lunar and Planetary Laboratory}, \orgname{University of Arizona}, \orgaddress{ \city{Tucson}, \state{AZ}, \country{USA}}}
\affil[13]{\orgdiv{Jet Propulsion Laboratory}, \orgname{California Institute of Technology}, \orgaddress{ \city{Pasadena}, \state{CA}, \country{USA}}}

\maketitle

\vspace{-5cm}

\textbf{Most stars, including our Sun, will one day evolve into red giants and, subsequently, white dwarfs. Several planet candidates have recently been identified orbiting white dwarfs\cite{Vanderburg2020,Blackman2021,Mullally2024,Limbach2024}, demonstrating that planets can survive the stellar post-main-sequence stage intact. Little is known about the atmospheric composition of post-main-sequence planets, with the most evolved transiting planets with atmospheric detections to date orbiting subgiants\cite{Kilpatrick2018,Colon2020}. Here we report an atmospheric detection for the white dwarf planet WD~1856~b, achieved through transmission spectroscopy with the JWST NIRSpec PRISM. Our 0.5--5.0\,$\mu$m spectrum reveals the presence of hydrocarbons (odds ratio of 167:1--5377:1, with CH$_4$ preferred at 17:1--30:1), aerosols (2×10$^5$:1--2×10$^6$:1), and thermal emission from the planetary nightside (2×10$^{63}$:1--2×10$^{73}$:1). Our spectral analysis constrains WD~1856~b's mass to 4.3--10.9\,$M_{\rm J}$, finds a carbon-enriched atmosphere (with a CH$_4$ abundance of $\approx$ 7\,\%), and an effective temperature exceeding the expected planetary equilibrium temperature (390--412\,K vs. 160\,K). Based on cooling models, these results suggest that WD~1856~b underwent a migration-related reheating event 3.0--5.5\,Gyr into the white dwarf phase, consistent with post-main-sequence tidal evolution to the present-day 0.02\,au circular orbit. Our results provide a window into the ultimate fate of giant planets orbiting stars with masses similar to our Sun.}


\maketitle


\maketitle

We observed a transit of WD~1856~b on 27 April 2023 with JWST’s Near InfraRed Spectrograph (NIRSpec), using the PRISM mode, as part of Guest Observer Program 2358 (PI: Ryan MacDonald). WD~1856~b is a cool ($T_{\rm{eq}} = 160$\,K) Jupiter-sized (0.9\,$R_J$) planet transiting the white dwarf WD~1856+534\cite{Vanderburg2020}. The host white dwarf ($T_{\rm{eff}} =$ 4710\,K, 0.0131\,$R_{\rm{Sun}}$, 0.518\,$M_{\rm{Sun}}$, $t_{\rm{cool}} \sim$ 6\,Gyr;\cite{Vanderburg2020}) is 25\,pc from Earth and has a DA spectral class\cite{Alonso2021}. The current close-in orbit of WD~1856~b (0.02\,au / 1.4-day) requires orbital evolution after the main sequence to avoid engulfment during the red giant phase. Hypotheses for WD~1856~b's orbital evolution include common-envelope evolution during the red giant or asymptotic giant branch (AGB) phase\cite{Lagos2021,Merlov2021} or, alternatively, high-eccentricity migration\cite{O'Connor2021}. Distinguishing these scenarios has proven challenging, given the uncertain planetary mass\cite{Vanderburg2020,Xu2021,Limbach2025}. As the host white dwarf WD~1856 is a relatively faint target (J = 15.677), we prioritized the wide wavelength range (0.6--5.5~$\mu$m at a mean resolving power of $\lambda/\Delta\lambda \sim 100$) and high throughput of the NIRSpec PRISM over other JWST instrument modes. Our observation lasted 1.98 hours, of which WD~1856~b's transit lasted 8 minutes.

\begin{figure*}
    \centering
    \includegraphics[width=\textwidth]{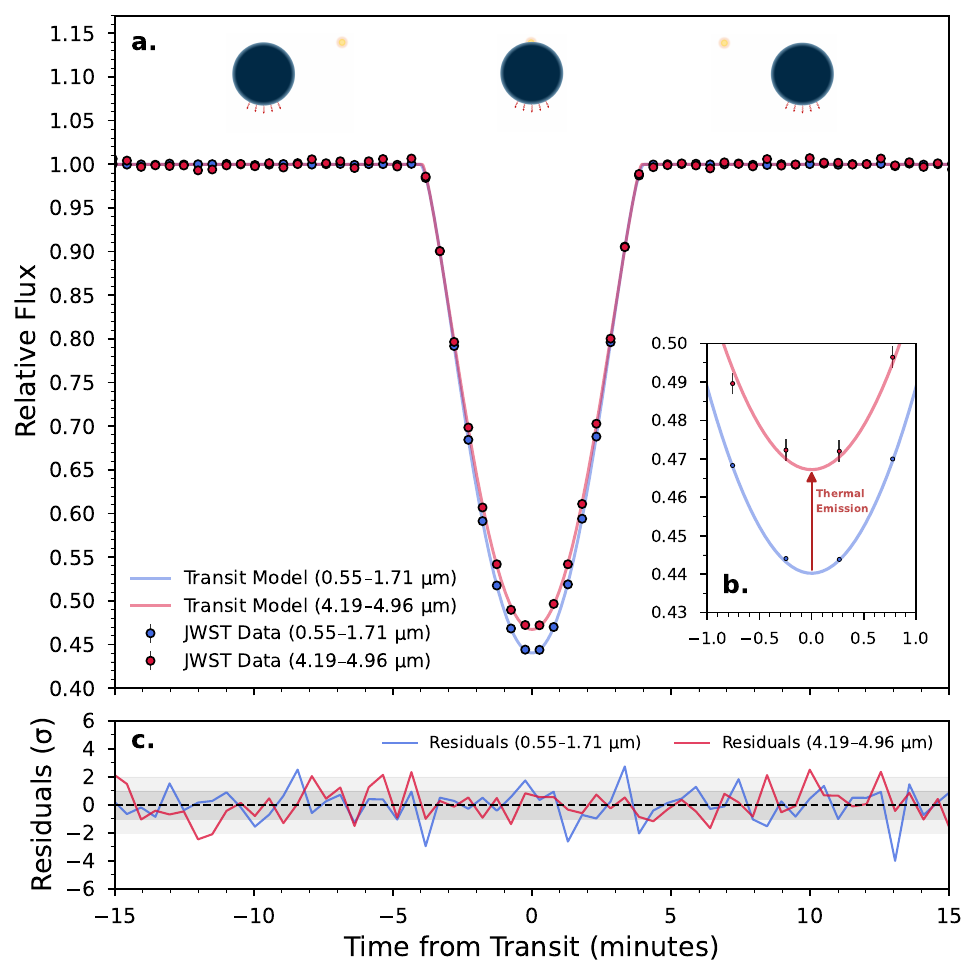}
    \caption{\textbf{Detection of nightside thermal emission from WD~1856~b's atmosphere.} \textbf{a.} JWST NIRSpec PRISM transit observations of the white dwarf planet WD~1856~b integrated over two broadband wavelength regions: 0.55--1.71\,$\mu$m (blue data points) and 4.19--4.96\,$\mu$m (red data points). The best-fitting model transit light curves are overplotted (blue and red curves). The transit event is shallower at longer wavelengths due to planetary thermal emission diluting the transit depth. The top schematic depicts the grazing transit geometry of the WD~1856 system, with the planet and white dwarf shown to scale. \textbf{b.} Zoom-in of the mid-transit to demonstrate the high signal-to-noise detection of the nightside thermal emission. The 1\,$\sigma$ errors for the blue data points are too small to be seen (the mean 1\,$\sigma$ errors are $5.3\times 10^{-4}$ from 0.55--1.71\,$\mu$m and $2.8\times 10^{-3}$ from 4.19--4.96\,$\mu$m). \textbf{c.} Residuals between the data and best-fitting model.}
    \label{fig:lightcurves}
\end{figure*}

\begin{figure*}
    \centering
    \includegraphics[width=\textwidth]{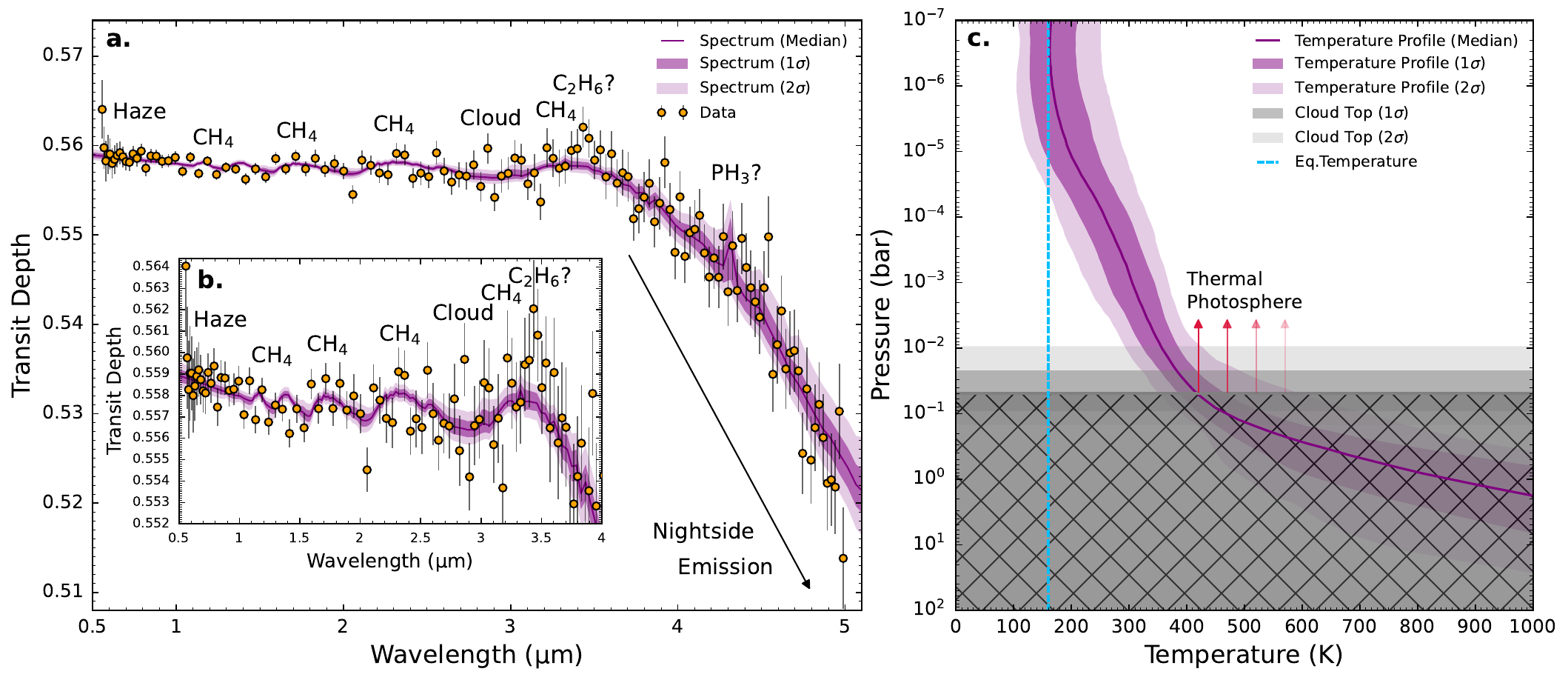}
    \caption{\textbf{Atmospheric retrieval of WD~1856~b's transmission spectrum.} \textbf{a.} WD~1856~b's JWST NIRSpec PRISM transmission spectrum (\firefly data reduction; orange circles with 1\,$\sigma$ error bars) is compared to the retrieved model spectrum (purple line and contours, showing the median, 1\,$\sigma$, and 2\,$\sigma$ credible intervals). Multiple absorption features from CH$_4$ are detected, alongside nightside thermal emission, continuum aerosol opacity, and tentative evidence of PH$_3$ and C$_2$H$_6$. \textbf{b.} Zoom-in to highlight the short-wavelength spectral features. \textbf{c.} Retrieved temperature profile (purple contours) and cloud-top pressure (grey contours) from WD~1856~b's transmission spectrum. An optically-thick cloud deck near 100\,mbar (grey hatching) blocks thermal emission from deeper layers. The top of the hatched region corresponds to the median retrieved cloud-top pressure, while the 1\,$\sigma$, and 2\,$\sigma$ credible regions appear above (horizontal grey shading). Most thermal emission arises from the cloud-top pressure near 100\,mbar (red arrows).}
    \label{fig:Retrieved_spectrum_PT}
\end{figure*}

We reduced the JWST observations using two independent data pipelines, 
\firefly\cite{Rustamkulov2022} and \Juniper (newly presented here, see Methods). Each reduction pipeline yielded sets of spectroscopic transit light curves for WD~1856~b, which we independently fit with a transit model (Methods). Since most existing stellar model grids do not include models for evolved or remnant objects, our transit models used a custom limb darkening prescription for WD~1856 derived from a white dwarf model\cite{Blouin2018} fitted to the out-of-transit host flux (Methods). Figure~\ref{fig:lightcurves} shows transit light curves of WD~1856~b from \firefly, integrated over two broadband wavelength regions from 0.55--1.71\,$\mu$m and 4.19--4.96\,$\mu$m. Our JWST observations find a 3\% shallower transit depth in the second, near-infrared, wavelength region compared to visible wavelengths. The physical mechanism for this transit depth difference is planetary thermal emission, as we establish below. Our PRISM transmission spectrum of WD~1856~b from the \firefly code is shown in Figure~\ref{fig:Retrieved_spectrum_PT} (left panel).

We model WD~1856~b's transmission spectrum using the radiative transfer and retrieval code \POSEIDON\cite{MacDonald2017, MacDonald2023}, adapted here to include partial planet-star overlap (i.e. grazing transit geometry)  and the effect of nightside thermal emission\cite{Kipping2010,Kappelmeier2024} (see Methods). We consider the main carbon, oxygen, nitrogen, sulphur, and phosphorous carriers expected in a H$_2$ dominated atmosphere at WD~1856~b's equilibrium temperature, and several disequilibrium tracers, resulting in the following gases: CH$_4$, NH$_3$, H$_2$O, CO$_2$, CO, C$_2$H$_2$, H$_2$S, PH$_3$, HCN, C$_2$H$_4$, and C$_2$H$_6$. Our retrievals freely fit the temperature structure, chemical composition, an opaque cloud deck, and a power law haze\cite{MacDonald2017}. Given WD~1856~b’s uncertain mass, our retrievals fit the mass jointly with the atmospheric properties. We conducted retrievals on both data reductions, finding excellent consistency between \firefly and \Juniper (see Methods). The retrieved spectrum and temperature structure for the \firefly reduction are shown in Figure~\ref{fig:Retrieved_spectrum_PT}.

\begin{figure*}
    \centering
    \includegraphics[width=\textwidth]{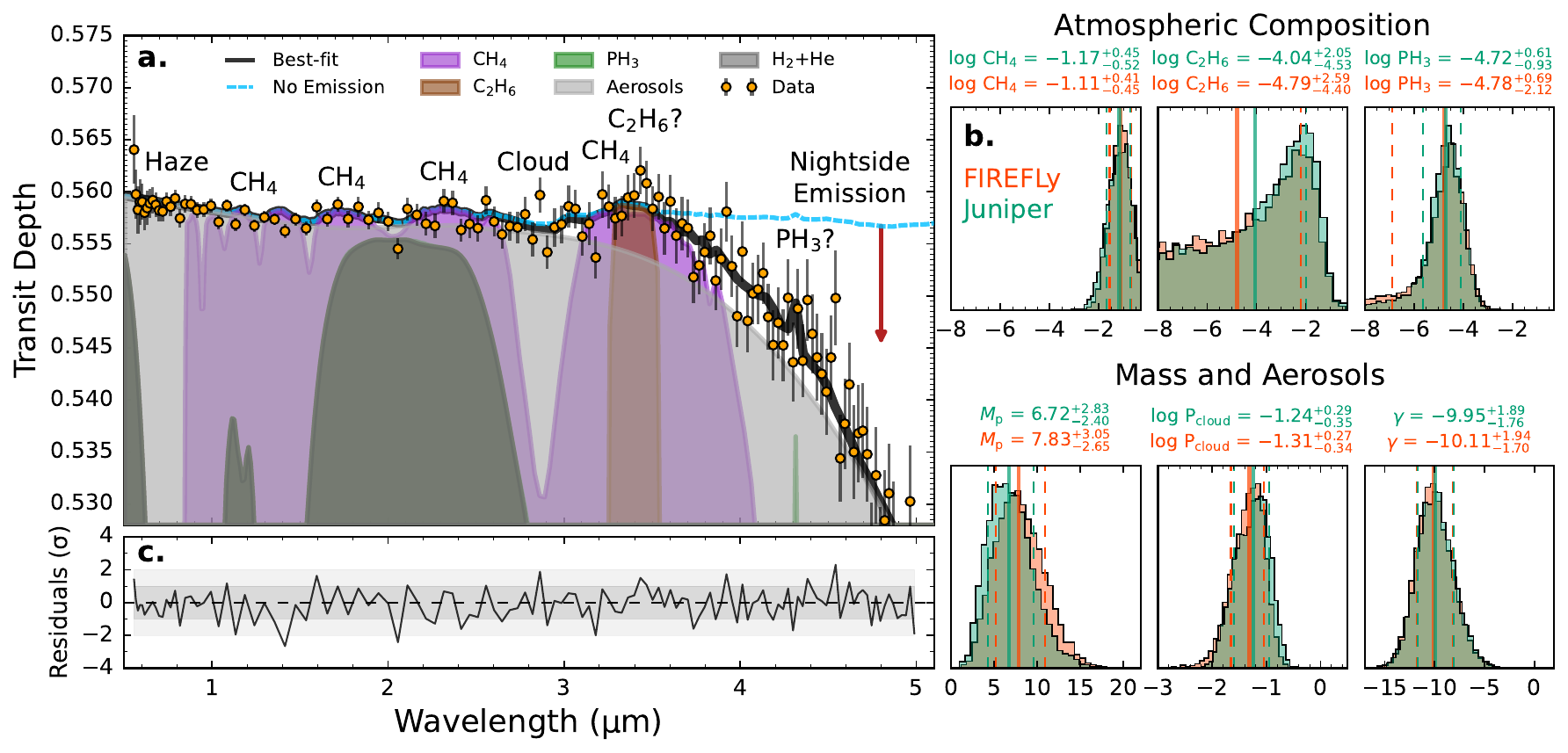}
    \caption{\textbf{Best-fitting spectrum and atmospheric properties of WD~1856~b.} \textbf{a.} Spectral decomposition illustrating the model components required to explain WD~1856~b's JWST transmission spectrum (\firefly reduction; orange circles with 1\,$\sigma$ error bars). The best-fitting model (black curve) contains absorption from multiple CH$_4$ features (purple shading), H$_2$ + He continuum absorption (dark grey shading), and tentative evidence of C$_2$H$_6$ near 3.4\,$\mu$m (brown shading) and PH$_3$ near 4.3\,$\mu$m (orange shading). Aerosol continuum opacity (silver shading) blocks thermal emission from the deep atmosphere in windows between CH$_4$ bands, obscuring the `negative spikes' that would otherwise be observed near 0.7\,$\mu$m, 2.9\,$\mu$m, and longwards of 4.0\,$\mu$m for a clear atmosphere. The best-fitting model without nightside thermal emission (blue dashed curve) cannot explain the significantly lower transit depths longwards of 3.5\,$\mu$m. \textbf{b.} Retrieved atmospheric properties, planetary mass, and aerosol properties from WD~1856~b's JWST transmission spectrum for the \firefly (orange histograms) and \Juniper (green histograms) data reductions. The median (solid line) and 1\,$\sigma$ credible intervals for each parameter (dashed lines) are overlaid. \textbf{c.} Residuals between the data and best-fitting model.}
    \label{fig:contribution_histograms}
\end{figure*}

Our retrievals establish at least one hydrocarbon (4.5\,$\sigma$ / $\ln \mathcal{B} = 8.59$ for \Juniper, 3.6\,$\sigma$ / $\ln \mathcal{B} = 5.12$ for \firefly) is present in WD~1856~b's atmosphere. The main contributor to the hydrocarbon inference is CH$_4$ (3.1\,$\sigma$ / $\ln \mathcal{B} = 3.40$ for \Juniper, 2.9\,$\sigma$ / $\ln \mathcal{B} = 2.87$ for \firefly), but C$_2$H$_6$ also contributes to the best-fitting model.  Figure~\ref{fig:contribution_histograms} (left panel) shows spectral contributions to our best-fitting model, which achieves an excellent fit to the \firefly data ($\chi^2_{\nu}$ = 1.03 with 102 degrees of freedom). The fit quality is slightly less good for \Juniper ($\chi^2_{\nu}$ = 1.28, indicating consistently with the data within 2$\sigma$ for 102 degrees of freedom), but the retrieval results are nonetheless consistent between the two data reductions. CH$_4$ causes three prominent absorption bands near 1.75\,$\mu$m, 2.3\,$\mu$m, and 3.3\,$\mu$m, with the retrieved CH$_4$ abundance (2--20\,\%) indicating an atmospheric metal enrichment of $\sim$ 100\,$\times$ solar (log C/H = $2.24^{+0.33}_{-0.40}$ for \firefly and $2.21^{+0.35}_{-0.44}$ for \Juniper in solar units, with corresponding 3\,$\sigma$ lower limits of 10.8$\times$ solar and 7.9$\times$ solar, respectively). We confirmed the CH$_4$ inference arises from multiple bands by repeating a \firefly retrieval with the data from 3.1--3.5\,$\mu$m masked, yielding a 3.0\,$\sigma$ model preference. The best-fitting model includes a contribution from PH$_3$ near 4.3\,$\mu$m, which could be an indicator of disequilibrium vertical mixing as in Jupiter's upper atmosphere\cite{Larson1977}. However, model degeneracies and lower signal-to-noise at longer wavelengths preclude a detection of PH$_3$ with the present data. Similarly, a feature attributable to C$_2$H$_6$ at 3.4\,$\mu$m is partially degenerate with CH$_4$ absorption near 3.3\,$\mu$m, so C$_2$H$_6$ is not significantly detected. However, the combined evidence for any hydrocarbon (i.e. CH$_4$, C$_2$H$_2$, C$_2$H$_4$, and C$_2$H$_6$) is statistically significant at $\approx 4\,\sigma$. No other chemical species (e.g. NH$_3$ or H$_2$O) are detected (see Methods, Extended Data Figure~\ref{fig:Retrieval_cornerplot} for upper limits).

Besides hydrocarbons, we detect thermal emission from the observer-facing nightside hemisphere (18.5\,$\sigma$ / $\ln \mathcal{B} = 169$ for \Juniper and 17.3\,$\sigma$ / $\ln \mathcal{B} = 146$ for \firefly) and aerosols (5.7\,$\sigma$ $\ln \mathcal{B} = 14.5$ for \Juniper and 5.3\,$\sigma$ / $\ln \mathcal{B} = 12.3$ for \firefly). Emission from the nightside hemisphere ($\sim$ 400\,K) causes the significant downwards slope longwards of 3.5\,$\mu$m. Since thermal emission in opacity windows arises from deeper, hotter layers, an optically thick cloud deck must exist for pressures deeper than $\sim$ 10\,mbar (Figure~\ref{fig:Retrieved_spectrum_PT}, right panel) to explain the lack of `negative spikes' between the CH$_4$ absorption bands (Figure~\ref{fig:contribution_histograms}, left panel). Since most thermal emission arises from the cloud-top temperature, the recent measurement of a $\sim$ 190\,K thermal excess in MIRI photometry of the WD~1856 system\cite{Limbach2025} (at a different epoch near quarter phase) potentially indicates variable cloud-top pressures for different hemispheres or in time. Alternatively, an unidentified systematic or calibration issue affecting either instrument may reconcile the difference, but we did not identify any such effect in either dataset. Additional evidence of aerosols arises from the scattering slope shortwards of 1\,$\mu$m, which is more prominent than H$_2$ Rayleigh scattering from a clear atmosphere. We investigated several specific aerosols using a Mie scattering model, but the present data are insufficient to identify the aerosol's chemical composition (see Methods). Finally, we report the first constrained measurement of WD~1856~b's mass: 4.3--10.9\,$M_{\rm J}$ (1\,$\sigma$ spread across both data reductions, or $6.7^{+2.8}_{-2.4}$\,$M_{\rm J}$ for \firefly and $7.8^{+3.1}_{-2.7}$\,$M_{\rm J}$ for \Juniper). Our mass constraint arises from a scale height trade between the CH$_4$ absorption features and the integrated optical depth to the emitting pressure of the nightside atmosphere.

The atmospheric temperature at which WD~1856~b radiates to space ($T_{\rm{eff}} =$ 390–412\,K, or $400^{+6}_{-10}$\,K for \firefly and $405^{+7}_{-11}$\,K for \Juniper; see Methods) is substantially elevated over both the equilibrium temperature (160\,K) and that expected for an evolved giant planet at the $\sim$ 10\,Gyr system age ($\lesssim$ 100\,K)\cite{Vanderburg2020}. Given the retrieved mass and the present-day circular orbit, internal power sources such as deuterium fusion or tidal heating cannot contribute to the observed effective temperature (see Methods). However, a thermal reset of WD~1856~b can have occurred by either tidal heating during high-eccentricity migration\cite{O'Connor2021} or immersion in the stellar envelope during a common-envelope phase\cite{Lagos2021, Merlov2021}. These two scenarios predict different migration times relative to the death of the host star: common-envelope evolution coincides with the end of the progenitor's AGB phase ($5.4 \pm 0.7$\,Gyr ago; see Methods) and lasts $\sim 1$\,Myr\cite{Glanz2018}, whereas high-eccentricity migration can occur anytime during the white dwarf phase. Because substellar objects cool down at a predictable rate, our planetary mass and temperature constraints provide critical information to infer the reheating epoch.

\begin{figure*}
    \centering
    \includegraphics[width=\textwidth]{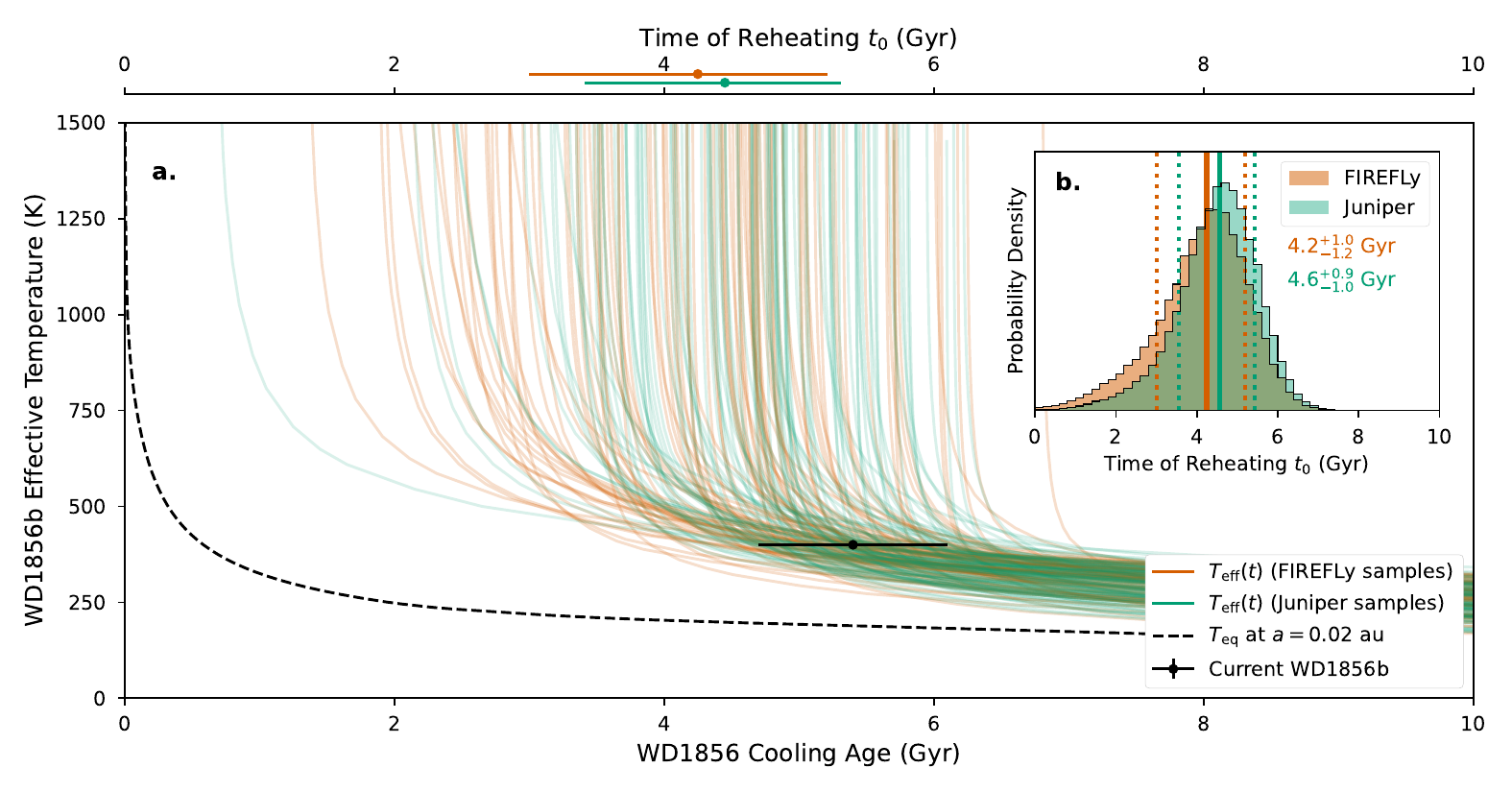}
    \caption{\textbf{Evolutionary history of WD~1856~b.} \textbf{a.} Reconstructed thermal evolution of WD~1856~b. An ensemble of randomly drawn thermal histories compatible with the JWST-derived constraints on WD~1856~b's mass, effective temperature, and the host white dwarf's cooling age are shown (orange curves for the \firefly data, green curves for \Juniper). The thermal history for an object at WD~1856~b's 0.02\,au orbit maintaining the zero-albedo equilibrium temperature is overlaid for comparison (dashed black curve). \textbf{b.} Inferred host cooling age at the time of WD~1856~b's reheating during migration, $t_0$, based on the ensemble of reconstructed thermal histories (orange and green histograms for \firefly and \Juniper). The median (solid line) and 1\,$\sigma$ credible interval (dashed lines) are overlaid and also plotted above panel a. Since common-envelope evolution predicts $t_0$ consistent with zero, the present-day mass and temperature favour migration billions of years after the conclusion of the asymptotic giant branch (AGB) phase.}
    \label{fig:thermal_evolution}
\end{figure*}

We reconstructed WD~1856~b's thermal evolution using theoretical cooling models for substellar objects (see Methods) to extrapolate its effective temperature backwards from the present. Figure~\ref{fig:thermal_evolution} shows a random set of reconstructed thermal histories for WD~1856~b, alongside the equilibrium temperature evolution (tracing the white dwarf's cooling). The range of possible histories is governed by uncertainties on WD~1856~b's mass and the host's cooling age. For each thermal history, we calculated the white dwarf's cooling age at the time of reheating (denoted $t_{0}$; see Methods). In Figure~\ref{fig:thermal_evolution}, $t_{0}$ corresponds roughly to the time when each cooling model intercepts the top of the diagram. We find WD~1856~b's reheating occurred 3.0--5.5\,Gyr after the end of the AGB phase ($4.2^{+1.0}_{-1.2}$\,Gyr for \firefly and $4.6^{+0.9}_{-1.0}$\,Gyr for \Juniper, see Figure~\ref{fig:thermal_evolution}, inset). More conservative  2\,$\sigma$ lower bounds on $t_{0}$ imply reheating occurred at least $1.4$\,Gyr (\firefly) or $2.1$\,Gyr (\Juniper) after the AGB phase. The AGB and post-AGB/pre-white-dwarf phases are extremely brief by comparison, lasting less than $2$\,Myr and $0.1$\,Myr, respectively (see Methods). The timing of the inferred reheating event is, therefore, inconsistent with common-envelope evolution during either phase. Therefore, WD~1856~b most likely underwent high-eccentricity migration to its current orbit, with the inferred reheating event corresponding to tidal circularisation.

WD~1856~b represents the first well-characterised transiting planet orbiting a white dwarf. The inferred thermal evolution of WD~1856~b demonstrates that high-eccentricity migration is a plausible fate for giant planets after the stellar main sequence. The retrieved CH$_4$ abundance is similar to Neptune's deep atmosphere (4\%\cite{Karkoschka2011}), which requires significant carbon enrichment of the planet's H$_2$ envelope from volatile-rich material, whether accreted before its migration\cite{Zahnle1994, Lellouch2002, Ginzburg2020} or afterwards\cite{Seligman2022}. This high atmospheric metallicity ($\gtrsim$ 100$\times$ solar) enhances aerosol production\cite{Visscher2010, Morley2012, Horst2018}, consistent with the detection of a short wavelength scattering slope in our transmission spectrum. As WD~1856~b demonstrates, spectroscopy of planets orbiting white dwarfs offers a new opportunity to determine the fate of planetary systems after the death of their star.

\clearpage

\newpage

\newpage

\newpage

\clearpage

\renewcommand{\figurename}{Extended Data Fig.}
\renewcommand{\tablename}{Extended Data Table}
\setcounter{figure}{0}
\setcounter{table}{0}

\clearpage

\section*{Methods} \label{methods}

\subsection*{Data reduction}

Our PRISM observation used 33 groups across 240 integrations, with each integration lasting 29.8 seconds, for a total observing time of 1.98 hours. While the transit duration of WD~1856~b only lasts 8 minutes, we selected this observing window to ensure JWST captured a transit of WD~1856~b with sufficient out-of-transit baseline for detector settling. We used two independent codes to reduce the NIRSpec PRISM observation of WD~1856~b and extract transmission spectra: \firefly and \Juniper. Here we detail the reduction and light curve fitting approach used by each code. 

\subsubsection*{\firefly}

We first reduced the WD~1856~b data using the Fast InfraRed Exoplanet Fitting Lyghtcurve (\firefly)\cite{Rustamkulov2022,Rustamkulov2023,Sing2024} reduction suite. The reduction started with the uncalibrated (\texttt{uncal.fits}) images and a customized \texttt{jwst} pipeline reduction. During the stage 1 and 2, the 1/f noise was removed at the group-level, using the top and bottom 6 rows to measure the count level for each column and subtracting the median value. The dark current step was skipped, and the jump step was performed with a rejection threshold of 20. During stage 3, we then used the custom-run pipeline 2D images after the \texttt{jwst.assign\_wcs} step, and performed customized cleaning of bad pixels, cosmic rays and hot pixels. We used cross-correlation to measure the positional shift of the spectral trace across the detector and shift-stabilized the images with flux-conserving interpolation. This procedure has been found to reduce the amplitude of position-dependent systematic trends\cite{Rustamkulov2022,Rustamkulov2023}. An aperture size of 5.7 pixels was used to extract the spectra, with this product used to fit the transit light curves and extract the exoplanet spectra.

During stages 4 and 5, we fit the white light curve using a linear baseline and a limb-darkened transit model\cite{Kreidberg2015}. The stellar limb-darkening was modeled with the procedures from Ref\cite{Sing2010} and a quadratic function using \texttt{ExoTiC-LD}\cite{Grant2022}. The values were fixed to the best-fitting theoretical host white dwarf model values (see the `White Dwarf Host Spectrum' section below). We measured the white light curve using the 0.55 to 3 $\mu$m region, such that the planetary emission would not bias the transit depth and resulting system parameters. The NIRSpec PRISM spectral time series for \firefly is shown in Extended Data Figure~\ref{fig:lightcurves}). Several pre-transit exposures showed abnormally low flux levels, which we flagged as outliers and removed from the remaining analysis. These outliers appear to be due to clusters of bright/hot pixels, so are likely associated with snowball events. We fixed the period using the results from Ref.\cite{Kubiak2023}. The best-fit white light curve system values are given in Extended Data Table~\ref{tab:data_fit_params}. The transmission spectral light curves used 3-pixel binning and were fit with the same model, setting the system parameters to be fixed to the white light curve values.

\subsubsection*{\Juniper}

We also applied \texttt{Juniper}, a new custom pipeline for JWST NIRSpec observations, to reduce our WD~1856~b data. \texttt{Juniper} contains a wrapper for Stages 1 and 2 of the \texttt{jwst} pipeline with custom steps. We start our processing from \texttt{Juniper} Stage 1 with the \texttt{uncal.fits} files from the Mikulski Archive for Space Telescopes (MAST). We opt to disable the \texttt{jwst} Stage 1 jump detection step and instead handle cosmic rays through custom procedures in later stages. Prior to ramp fitting, group-level background subtraction is performed using the top and bottom six rows as the background region to reduce scatter in the extracted light curves. We spatially filter $3\,\sigma$ outliers from this region and average along columns to determine the background level of counts per column, which is subtracted from the full group. We then proceed with \texttt{jwst} Stage 1 ramp fitting and gain scaling. \texttt{Juniper} Stage 2 is a pure wrapper for \texttt{jwst} Stage 2; we carry out this stage with the flat field and photom steps disabled, as neither is required to measure transit depth and we observed the former to increase the noise in the spectral extraction.

\texttt{Juniper} Stage 3 performs additional cleaning at the integration level. We first mask pixels flagged by the \texttt{jwst} pipeline for data quality issues. We then reject cosmic rays in time over two iterations, replacing $6.5\,\sigma$ outliers with the median value of the pixel in time. Finally, a second round of background subtraction, using the same strategy as the group-level background subtraction in Stage 1, is performed using the top and bottom three rows with outliers masked at $3\,\sigma$.

Stage 4 of the \texttt{Juniper} pipeline extracts 1D spectra, which are subsequently binned to produce light curves. Our aperture is centered on the brightest row of the trace and extends $\pm3$ pixels above and below it. We perform optimal extraction\cite{Horne1986} to extract the 1D spectrum, taking as our extraction profile the median image of the trace contained in the aperture, normalized along columns. We sum across all pixels from 0.552 to 3 $\mu$m to extract a broadband light curve; we choose not to include light from wavelengths longer than 3\,$\mu$m as this light is strongly affected by contamination from nightside thermal emission, which affects the determined system parameters (e.g. semimajor axis, transit epoch, impact parameter). We then bin every 3 pixels to produce 137 median-normalized spectroscopic light curves at nearly native resolution, spanning 0.552 to 5.360\,$\mu$m. Our extracted broadband and spectroscopic light curves would typically be sigma-clipped to further remove outliers; however, this technique is prone to clipping out the transit itself due to the short transit duration and large transit depth. We therefore disable this procedure and employ alternative outlier rejection procedures in Stage 5.

\texttt{Juniper} Stage 5 is the final stage of the pipeline which fits transit models to each light curve to extract transit depth and produce a transmission spectrum. Our fitting procedure is a two-step process combining linear and nonlinear fitting techniques, which we use to clean outliers that sigma-clipping cannot safely remove. We start by using a linear least-squares estimator to fit a \texttt{batman} transit model\cite{Kreidberg2015}, applying a quadratic limb darkening law with coefficients generated with \texttt{ExoTiC-LD}\cite{Grant2022} using a custom white dwarf model produced by fitting a white dwarf spectrum to the out-of-transit flux of WD~1856 (described below). Our model is further multiplied by a linear-in-time trend to account for visit-long ramp effects. We then compute the residuals of the fitted transit and systematics model and clip any points in the light curve that produce $3\,\sigma$ outliers in the residuals. We compute the standard deviation of the sigma-clipped residuals to estimate the photometric uncertainty, which we supply to the next step of our fit procedure. We re-fit the sigma-clipped light curve using Markov Chain Monte Carlo methods\cite{ForemanMackey2013} to extract our final planet-star radius ratio spectrum. We first fit our 0.552-3.0\,$\mu$m broadband light curve using this two-step fitting process to determine the broadband depth, semimajor axis $a/R^*$, inclination $i$, mid-transit epoch $t_C$, and linear-in-time systematics model parameters, from which we derive the impact parameter $b$ and its uncertainty. The orbit period $P$ was held fixed to 1.407939217~d based on a follow-up paper studying transit timing variations in the WD 1856+534 system\cite{Kubiak2023}. Our broadband light curve analysis yielded $a/R^*=339.25\pm5.92$ and $b=7.34\pm0.20$. We then fix these values as well as $t_C$ as determined by the broadband light curve fit for all subsequent spectroscopic light curve fits. We fit every spectroscopic light curve with our two-step process to determine $R_p(\lambda)/R_*$ and the linear systematics trend in every wavelength channel. Our broadband light curve fit achieves residuals of 522 ppm, while our spectroscopic fits achieve median residuals of 8153 ppm. We present our fitted system parameters ($a/R_*$, $b$, $R_p/R_*$) and broadband transit depth in Extended Data Table~\ref{tab:data_fit_params}.

\subsection*{Grazing Transit Spectroscopy}

The unique transit geometry of the WD~1856 system required us to develop a novel approach to express and model transmission spectra. A transmission spectrum encodes the wavelength-dependent effective area of a planet relative to its host star. Exoplanet analyses typically take the spectroscopic planet-star radius ratio from light curve fits, $R_p (\lambda) / R_*$, then express the transmission spectrum as $R_p (\lambda)^2/R_*^2$. This quantity is equivalent to the transit depth for a planet with radius $R_p (\lambda)$ fully occulting a non-limb-darkened star of radius $R_*$. However, since WD~1856~b is is 7$\times$ larger than its white dwarf host with a grazing transit, the transmission spectrum cannot be written as $R_p (\lambda)^2/R_*^2$. Indeed, WD~1856~b's transit depth is time-dependent throughout the transit (see\cite{Xu2021}) with a maximum transit depth corresponding to the instant of greatest areal overlap between the planet and its host (Figure~\ref{fig:lightcurves}). We therefore express WD~1856~b's transmission spectrum as the wavelength-dependent \emph{maximum transit depth} at the time of mid-transit, $A_{\rm{p}}/A_*$.

We convert the spectroscopic planet-host radius ratio into the mid-transit transit depth by calculating the time-dependent area overlap of two discs. The overlapping area of two circles with radii $R_p$ (representing the planet) and $R_*$ (representing the white dwarf), separated by a distance $d$, is given by:

\begin{equation}
    A_{\rm{p}} (d) = R_{\rm{p}}^2 \theta + R_{\rm{*}}^2 \phi - \frac{1}{2} R_{\rm{p}}^2 \sin(2\theta) - \frac{1}{2} R_{\rm{*}}^2 \sin(2\phi) 
\label{eq:A_p_overlap}
\end{equation}
where
\begin{equation}
    \theta = \cos^{-1} \left(\frac{d^2 + R_{\rm{p}}^2 - R_{\rm{*}}^2}{2 \, d \, R_{\rm{p}}}\right)
\label{eq:theta}
\end{equation}
\begin{equation}
    \phi = \cos^{-1} \left(\frac{d^2 + R_{\rm{*}}^2 - R_{\rm{p}}^2}{2 \, d \, R_{\rm{*}}}\right)
\label{eq:phi}
\end{equation}

\noindent Considering the time of mid-transit (when $d = b \, R_*$, where $b$ is the transit impact parameter), we can express the maximum transit depth as:

\begin{equation}
    \frac{A_{\rm{p}}}{A_{\rm{*}}} = \frac{1}{\pi} \left[ \left( \frac{R_p}{R_*} \right)^2 \left( \theta - \frac{1}{2} \sin 2\theta \right) + \left( \phi - \frac{1}{2} \sin 2\phi \right) \right]
    \label{eq:ApAs_mid-transit}
\end{equation}
where
\begin{equation}
    \theta = \cos^{-1} \left[ \frac{b^2 + \left( \frac{R_p}{R_*} \right)^2 - 1}{2b \left( \frac{R_p}{R_*} \right)} \right]
\label{eq:theta_mid-transit}
\end{equation}
\begin{equation}
    \phi = \cos^{-1} \left[ \frac{b^2 - \left( \frac{R_p}{R_*} \right)^2 + 1}{2b} \right]
\label{eq:phi_mid-transit}
\end{equation}

We use Equations~\ref{eq:ApAs_mid-transit}, ~\ref{eq:theta_mid-transit}, and ~\ref{eq:phi_mid-transit} to map the spectroscopic radius ratio, $R_p/R_*$, and impact parameter from each data reduction's spectroscopic light curve fits into the equivalent mid-transit transmission spectrum. We use the \texttt{uncertainties} Python package to propagate errors using these formulae. This approach automatically removes offsets between the different reductions for $R_p/R_*$, since each corresponding pair of $R_p/R_*$ and $b$ must yield consistent $A_p/A_*$ to have the same transit shape (i.e. to match Figure~\ref{fig:lightcurves}). 

\begin{figure*}
    \centering
    \includegraphics[width=\textwidth]{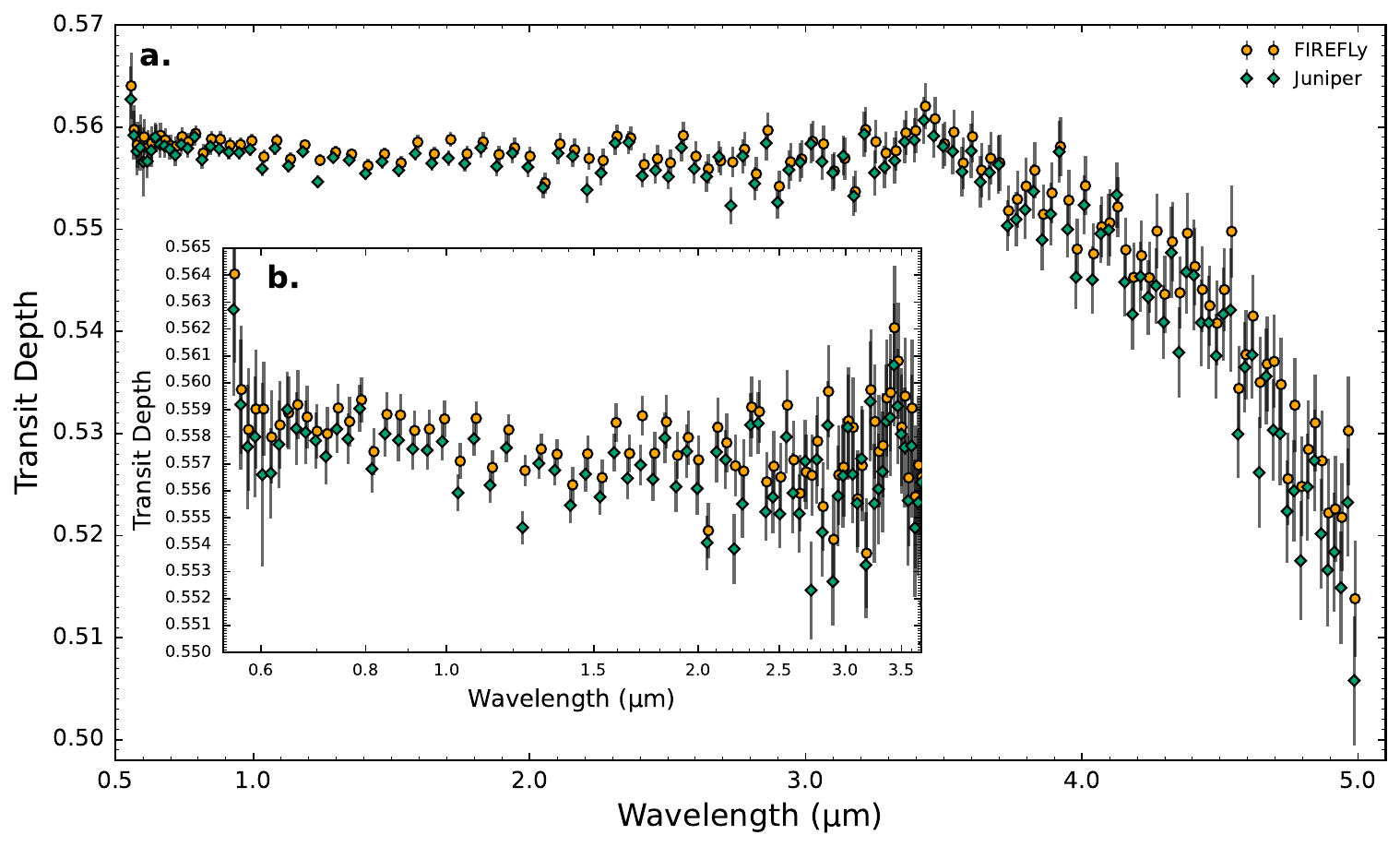}
    \caption{\textbf{Comparison of transmission spectrum data reductions.} \textbf{a.} The WD~1856~b transmission spectrum, expressed as the ratio between the mid-transit planet area occulting the host and the host surface area ($A_p / A_*$), is shown for FIREFLy (gold circles with 1\,$\sigma$ error bars) and Juniper (green diamonds with 1\,$\sigma$ error bars). \textbf{b.} Zoom-in on shorter wavelengths.}
    \label{fig:reduction_comparison}
\end{figure*}

We show our final transmission spectra of WD~1856~b, expressed as the mid-transit transmission spectrum ($A_p/A_*$), in Extended Data Figure~\ref{fig:reduction_comparison}. Both reductions clearly detect the strong signature of nightside contamination (the slope to lower transit depths at longer wavelengths) and lead to consistent atmospheric inferences from our retrieval analysis (see `\emph{Atmospheric Retrieval Analysis}'). We note that the two reductions partially deviate at wavelengths longer than 5\,$\mu$m --- largely due to differences in the light-darkening treatments and uncertainties in the red edge detector behaviour --- so we restricted our atmospheric analysis for WD~1856~b to the NIRSpec PRISM data from 0.5--5\,$\mu$m.

\subsection*{White Dwarf Host Spectrum}

\begin{figure*}
    \centering
    \includegraphics[width=\textwidth]{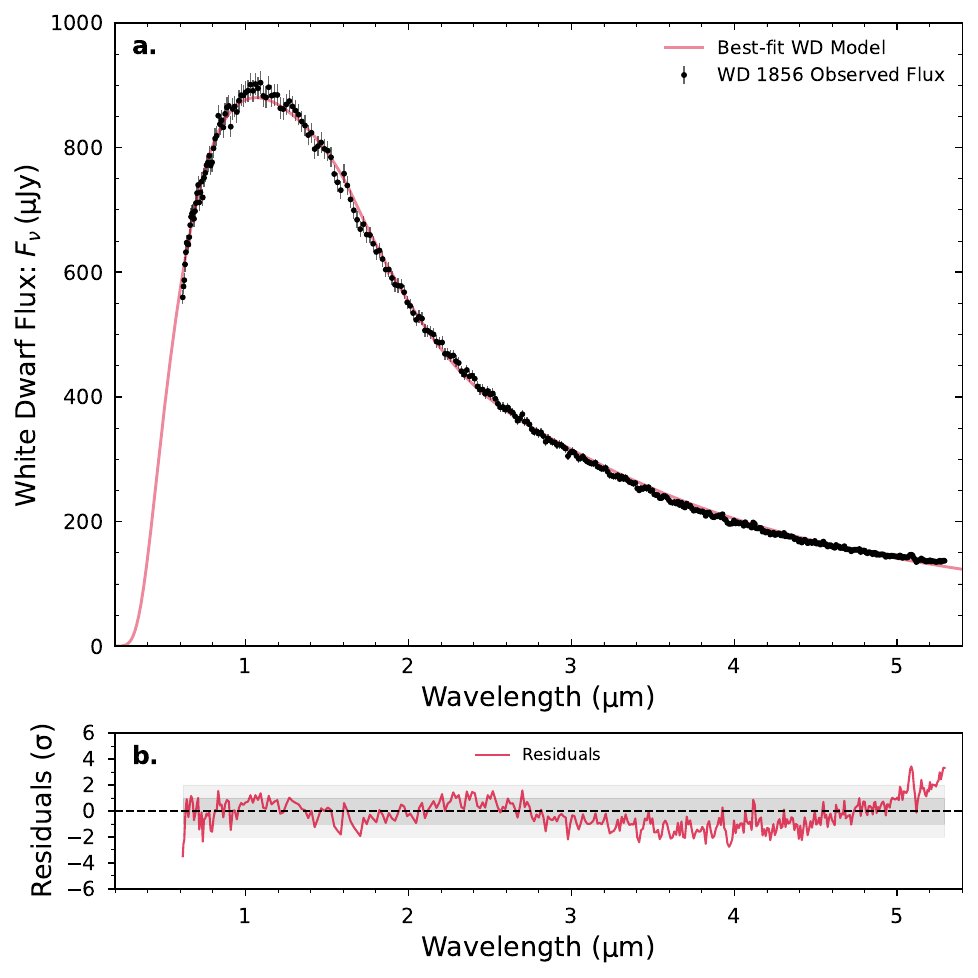}
    \caption{\textbf{WD~1856 host white dwarf spectrum.} \textbf{a.} JWST NIRSpec PRISM out-of-transit spectrum of WD~1856b (black) compared to a best-fitting white dwarf model (red). The 1\,$\sigma$ error bars include a minimum 2\% absolute flux calibration uncertainty. \textbf{b.} Residuals between the data and best-fitting model.}
    \label{fig:WD_host_spectrum}
\end{figure*}

We extracted a calibrated out-of-transit NIRSpec PRISM stellar spectrum for WD~1856 using the \firefly data reduction. Starting from the cleaned 2D images, we further flat fielded, flux-calibrated, and extracted the resulting host spectrum. The resulting stellar spectra is shown in Extended Data Figure~\ref{fig:WD_host_spectrum}.

We determined the host white dwarf's atmospheric parameters by fitting the out-of-transit system flux. We minimized the $\chi^2$ for a model suitable for cool white dwarfs\cite{Blouin2018} defined by three parameters: the white dwarf's effective temperature, $T_{\rm eff}$, its photospheric hydrogen-to-helium abundance ratio, and the solid angle $\pi {R_{*}}^2/D^2$. Since the distance $D$ is known from the \textit{Gaia} DR3 parallax, the solid angle directly constrains the white dwarf's radius. The radius, in turn, determines the white dwarf's mass and surface gravity given theoretical white dwarf structure models\cite{Bedard2020}. The best-fit solution (Extended Data Figure~\ref{fig:WD_host_spectrum}) corresponds to $T_{\rm eff}=4920\,$K, $\log g = 8.05$, and $N_{\rm H}/N_{\rm He}=4.1$. This solution yields an H$\alpha$ line that extends 2\% below the continuum, which is consistent with previously obtained optical spectroscopy\cite{Xu2021} (not considered here in our fit). We also attempted to fit the PRISM spectrum using pure-hydrogen models (i.e., without considering $N_{\rm H}/N_{\rm He}$ as a free parameter), but the best-fit solution yields a significantly worse fit to the PRISM data than the mixed H and He atmosphere solution. We calculated limb darkening coefficients for the best-fitting white dwarf model (using the approach from\cite{Gianninas2013}), which we then fixed during the WD~1856~b spectroscopic light curve fits for our two data reductions.

\subsection*{Transmission Spectrum Modeling} 

The grazing transit geometry of WD~1856~b, coupled with the clear presence of planetary nightside thermal emission, requires a novel modelling approach. The transmission spectrum for a planet with a grazing transit and nightside thermal emission can be written as\cite{MacDonald2022}:
\begin{equation}
    \Delta_{\lambda} = \frac{ A_{\rm{p \, (top)}} - \displaystyle\int_{A_{\rm{p}}} \mathcal{T}_{\lambda} \, dA }{\pi \, R_{*}^2} 
    \left( \frac{1}{1 + \displaystyle\frac{F_{\mathrm{p\, (night)}, \, \lambda}}{F_{*, \, \lambda}}} \right)
\label{eq:transmission_spectrum}
\end{equation}
where $A_{\rm{p \, (top)}}$ is the area of the planet overlapping the star at the top of the modelled atmosphere (given by Equation~\ref{eq:A_p_overlap} with $R_p = R_{p, \, \mathrm{top}}$), $\mathcal{T}_{\lambda}$ is the atmospheric transmissivity ($e^{-\tau_{\lambda}}$, where $\tau_{\lambda}$ is the slant optical depth) in the area element $dA$, and $F_{\mathrm{p\, (night)}}$ and $F_{*, \, \lambda}$ are the observed fluxes from the planetary nightside and white dwarf at Earth, respectively. We reduce the area integral in the first term to a single integral over the fractional annuli of the planet overlapping the star (i.e. $dA = A_p (r_{i, \, \mathrm{up}}) - A_p (r_{i, \, \mathrm{low}})$, where we use the radii of the upper and lower boundaries of each atmospheric layer in place of $R_p$ in Equation~\ref{eq:A_p_overlap}). The first term in Equation~\ref{eq:transmission_spectrum} represents the wavelength-dependent effective area of the fraction of the planet overlapping the white dwarf, relative to the white dwarf's projected disk area. The second term accounts for the `nightside pollution' / dilution of the transit depth\cite{Kipping2010} due to thermal emission from the planetary hemisphere facing the observer. 

Transmission spectra of WD~1856~b can also be expressed in terms of emergent fluxes by using the solid angle relation between the observed flux and the emergent (surface) flux, such that Equation~\ref{eq:transmission_spectrum} becomes:
\begin{equation}
\begin{aligned}
    \Delta_{\lambda} & = \frac{ A_{\rm{p \, (top)}} - \displaystyle\int_{A_{\rm{p}}} \mathcal{T}_{\lambda} \, dA }{\pi \, R_{*}^2}
    \\
    & \quad \times \left( \frac{1}{1 + \displaystyle\frac{R_{\mathrm{p, \, (night)}, \, \lambda}^2}{R_{*}^2} \frac{F_{\mathrm{p\, (night), \, surf}, \, \lambda}}{F_{*, \, \mathrm{surf}, \, \lambda}}} \right)
\end{aligned}
\label{eq:transmission_spectrum_2}
\end{equation}
where $R_{\mathrm{p, \, (night)}, \, \lambda}$ is the radius of the emitting thermal photosphere on the nightside (nominally the $\tau_{v, \lambda} = 2/3$ pressure level, where $\tau_{v, \lambda}$ is the vertical optical depth integrated downwards from the top of the atmosphere). The planet-star surface flux ratio featured in Equation~\ref{eq:transmission_spectrum_2} is a standard output from radiative transfer codes used to calculate exoplanet emission spectra. Similarly, the transmissivity, $ \mathcal{T}_{\lambda}$, is already calculated by radiative transfer codes calculating standard transmission spectra. Therefore, to calculate transmission spectra of WD~1856~b one can construct a model atmosphere and then calculate both the transmissivity from the slant optical depth and the emergent planet-host flux ratio. The observed transmission spectrum then represents a product between a grazing transit transmission spectrum and an `upside down' emission spectrum.

\subsection*{Atmospheric Retrieval Analysis}

We infer WD~1856~b's atmospheric properties via the open source Bayesian atmospheric retrieval code \POSEIDON\cite{MacDonald2017,MacDonald2023}. We model WD~1856~b's atmosphere using 100 layers spaced uniformly in log-pressure from $10^{-7}$--100\,bar. We assume the atmosphere is well-mixed, with consistent atmospheric properties at the day-night terminator and at the nightside, such that only a single set of parameters describe the atmospheric state. We fit for the planetary radius at the 10\,bar pressure level and the planetary mass, while fixing the white dwarf radius to $R_* = 0.0131 R_{\rm{Sun}}$ and the transit impact parameter to $7.430234$ (since the impact parameter uncertainty is already marginalized into the $R_p/R_*$ uncertainties, it does not need to be an independent free parameter). We include the $\log_{10}$ mixing ratios of the following molecules as free parameters: CH$_4$, NH$_3$, H$_2$O, CO$_2$, CO, HCN, C$_2$H$_2$, C$_2$H$_4$, C$_2$H$_6$,  H$_2$S, and PH$_3$. The remainder of the atmosphere is composed of H$_2$ and He with an abundance ratio of He/H$_2$ = 0.17, consistent with the giant planets in the solar system\cite{Atreya2020}. We parameterise WD~1856~b's temperature profile using an adaptation of a prescription used for brown dwarfs\cite{Piette2020}. This prescription retrieves the temperature at 9 pressure nodes (spaced uniformly per decade in pressure from 10$^-{6}$\,bar to 100\,bar) and interpolates between them with a spline. The 9 free parameters defining this temperature profile are the 100\,mbar temperature and 8 $\Delta T_{i}$ parameters encoding the temperature difference between each pair of nodes. Given the low external irradiation of WD~1856~b, we restrict $\Delta T_{i} > 0$ to consider physically plausible profiles with temperature monotonically increasing with pressure. Finally, we fit for a three-parameter aerosol model consisting of a power law scattering slope (with exponent $\gamma$) and an optically thick cloud-top pressure\cite{MacDonald2017}. We do not consider inhomogenous clouds around the terminator, since only a small fraction of WD~1856~b's terminator occults the white dwarf's surface during transit. 

Our retrieval model is thus defined by 25 free parameters, which we fit using MultiNest's\cite{Feroz2008,Feroz2009,Feroz2019} Python wrapper PyMultiNest\cite{Buchner2014} with 1,000 live points. The priors for each parameter are summarised in Extended data Table~\ref{tab:retrieval_priors}. We calculate Bayes factors (i.e. odds ratios; $\mathcal{B}$) via Bayesian model comparisons between nested retrieval models, with the retrieval model statistics summarised in Extended Data Table~\ref{tab:retrieval_model_stats}. For consistency with the exoplanet literature, we also convert the Bayes factor between two nested models (e.g. our reference model and a model excluding CH$_4$) into an `equivalent detection significance', $N\,\sigma$, using a standard relation\cite{Benneke2013}. We note, however, that there are several caveats associated with Bayes factor to detection significance mapping, so our preferred statistic for model preference is the Bayes factor/odds ratio (see Ref\cite{Kipping2025}).

We calculate model transmission spectra of WD~1856~b by solving Equation~\ref{eq:transmission_spectrum_2} on a wavelength grid ranging from 0.5--5.6\,$\mu$m at $R =$ 20,000. We sample high-resolution pre-computed cross sections\cite{MacDonald2022} onto this wavelength grid, using the following line list sources: CH$_4$\cite{Yurchenko2024}, NH$_3$\cite{Coles2019} H$_2$O\cite{Polyansky2018}, CO$_2$\cite{Yurchenko2020}, CO\cite{Li2015}, HCN\cite{Barber2014}, C$_2$H$_2$\cite{Chubb2020}, C$_2$H$_4$\cite{Gordon2022}, C$_2$H$_6$\cite{Gordon2022}, H$_2$S\cite{Azzam2016}, and PH$_3$\cite{Sousa-Silva2015}. We additionally include continuum opacity from H$_2$ and He collision-induced absorption\cite{Karman2019} and Rayleigh scattering. For the host flux, we use the best-fit white dwarf model shown in Extended Data Figure~\ref{fig:WD_host_spectrum}. Our model transmission spectra are finally convolved with the NIRSpec PRISM point spread function and binned down to the resolution of the observations to calculate the likelihood of each location in the retrieval model parameter space. 

\begin{figure*}
    \centering
    \includegraphics[width=\textwidth]{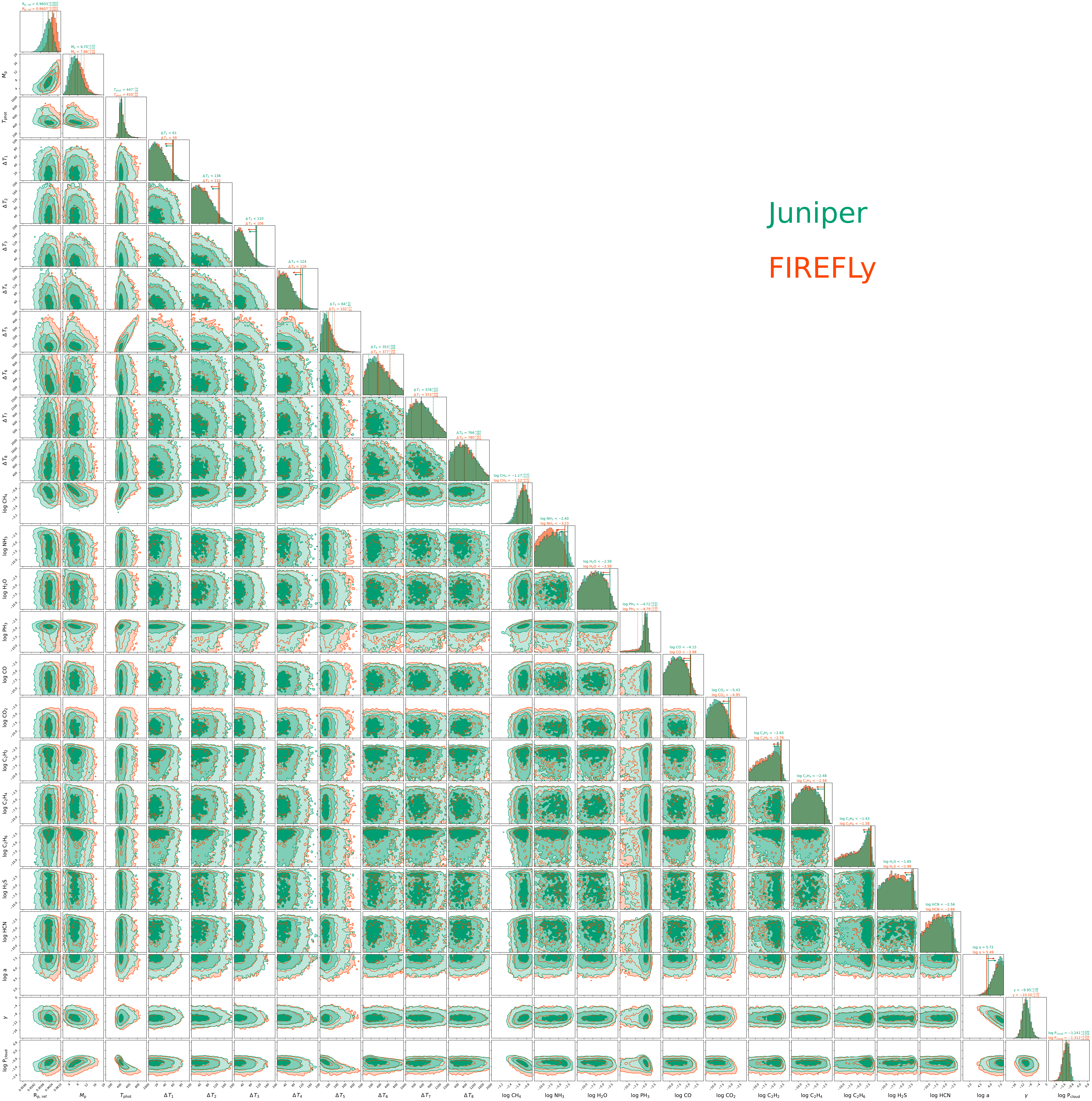}
    \caption{\textbf{Full posterior distributions for the \firefly (orange) and \Juniper (green) data reductions.} The histograms show the margianalised posterior probability density for each model parameter for each data reduction, with the median (solid lines) and 1\,$\sigma$ credible interval (dashed lines) overplotted for well-constrained parameters. Parameters without clear upper and lower limits, such as non-detected molecules, have 2\,$\sigma$ annotated instead (arrows). The 2D correlation panels are shaded according to the 1\,$\sigma$ (darkest) through to the 3\,$\sigma$ credible region.}
    \label{fig:Retrieval_cornerplot}
\end{figure*}

While Figure~\ref{fig:contribution_histograms} compares several retrieved atmospheric properties between the \firefly and \Juniper data reductions, we provide the full posterior distributions in Extended Data Figure~\ref{fig:Retrieval_cornerplot}. We find excellent agreement between \firefly and \Juniper for all retrieved parameters.

\begin{figure*}
    \centering
    \includegraphics[width=\textwidth]{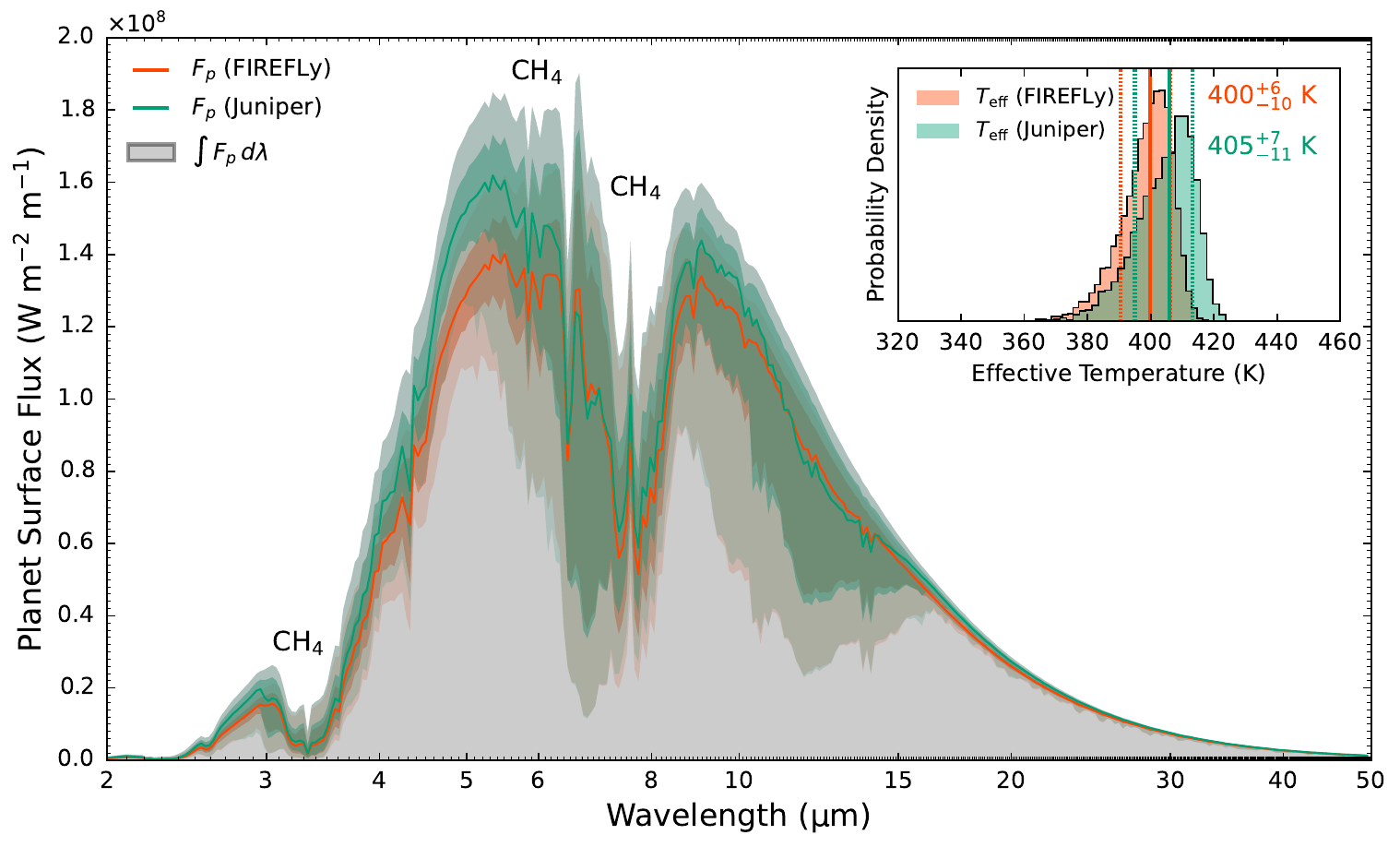}
    \caption{\textbf{Retrieved emergent flux and effective temperature.} \textbf{a.} WD~1856~b's retrieved emergent surface flux for the \firefly (orange curve and credible interval shading) and \Juniper (green curve and credible interval shading) reductions are shown extrapolated out to 50\,$\mu$m. The integral under the curve (grey shading) is used to determine the effective temperature, $T_{\rm{eff}}$, corresponding to the surface flux of each model atmosphere. \textbf{b.} Posterior distributions of WD~1856~b's effective temperature for the \firefly and \Juniper datasets, with the 1\,$\sigma$ credible region annotated.}
    \label{fig:Flux_Teff}
\end{figure*}

To interpret WD~1856~b's thermal history, we additionally calculate posterior distributions for the planetary effective temperature from our retrieval results. We calculated the emergent planetary surface flux of WD~1856~b for each atmosphere in the full set of posterior samples from both the \firefly and \Juniper reductions on a wavelength grid from 1--50\,$\mu$m. For each set of atmospheric parameters, we calculate the corresponding effective temperature using the Stefan-Boltzmann law: $T_{\mathrm{eff}} = \left( \frac{1}{\sigma_{SB}} \int \, F_{\mathrm{p, \, surf}, \, \lambda} \, d\lambda \right)^{1/4}$. Extended data Figure~\ref{fig:Flux_Teff} shows our retrieved surface flux spectrum for WD~1856~b for both data reductions and the corresponding $T_{\mathrm{eff}}$ posterior distributions. Adopting the lowest $1\sigma$ credible interval (from \firefly) and the highest $1\,\sigma$ credible interval (from \Juniper), we find a range of 390--412\,K for $T_{\mathrm{eff}}$. We similarly report the $1\,\sigma$ range encompassing both data reductions for $M_{\rm{p}}$ in the Main text. The $\approx$ 10\,K 1\,$\sigma$ uncertainty on $T_{\mathrm{eff}}$ is driven by the multiple CH$_4$ bands detected in our NIRSpec PRISM data setting the relative amplitude of other CH$_4$ features at longer wavelengths. However, the potential presence of other hydrocarbons, such as C$_2$H$_6$, allows a larger surface flux uncertainty where these species absorb in the mid-infrared (e.g. 10--15\,$\mu$m), which increases the uncertainty in the integrated power and hence $T_{\mathrm{eff}}$. Longer wavelength observations of WD~1856~b with MIRI LRS/MRS, such as those planned in JWST Cycle 4 (GO-9033 and GO-9157), will constrain $T_{\mathrm{eff}}$ even further.

\subsubsection*{Mie Scattering Retrievals}

We have established that models including aerosol opacity are required to explain WD~1856~b's transmission spectrum. Specifically, our free retrieval analysis above infers an opaque cloud deck near 100\,mbar and a haze to explain the power-law scattering slope shortwards of 1~$\mu$m. The enhanced scattering slope indicates a collection of small particles in the upper atmosphere, but our parametric description is agnostic to the specific aerosol composition. Here we consider retrievals including Mie scattering to investigate which specific aerosol species are consistent with WD~1856~b's transmission spectrum. 

The composition of small, Mie-scattering particles can be potentially identified via aerosol absorption features at infrared wavelengths, while their particle size is encoded by the scattering slope. We assess here which aerosol species and particle sizes can explain the observed scattering slope via retrievals including compositionally specific Mie-scattering aerosols. We do not test directly for specific species causing the opaque cloud deck, as this deck is likely composed of large particles with muted resonance features\cite{Marley1999, Min2004}. Since such a condensate cloud deck has no spectroscopic features, it is not possible to determine the composition unless condensates are lofted above the deck and become smaller in size. 

We use the Mie scattering retrieval module and database introduced in \POSEIDON v1.2\cite{Mullens2024}. Our Mie scattering retrievals use aerosol extinction cross sections pre-computed from refractive indices. We primarily consider a simple aerosol model parameterized by the $\log_{10}$ mean particle size (log $r_m$ $\mathcal{U}$[-3, 1]) and aerosol $\log_{10}$ volume mixing ratio (log aerosol $\mathcal{U}$[-30, -1]) --- representing a well-mixed aerosol uniformly distributed within the atmosphere. We also tested more complex aerosol models that fit for pressure-dependent aerosol mixing ratios, but these all reduced to a pressure-independent model. Our Mie retrievals use a six-parameter pressure-temperature (P-T) profile\cite{Madhusudhan2009}. We conduct these Mie retrievals on the \firefly data reduction.

We ran retrievals with a suite of aerosol species representing three different aerosol formation regimes that could be relevant in WD~1856~b's upper atmosphere. The first aerosol regime represents disequilibrium hazes and soot species that can be produced by photochemistry: titan tholins (Tholins;\cite{Khare1984, Ramirez2002}), carbon soot (C;\cite{Draine2003}), water-rich organic haze at two temperatures (ExoHaze 300K, ExoHaze 400K;\cite{He2023}), and hexene (C$_6$H$_{12}$;\cite{Anderson2000}). The second aerosol regime represents the myriad of sulfide and chloride clouds that form in brown dwarfs at the T-Y transition (400--1300\,K)\cite{Morley2012}, alongside Cr: chromium (Cr; Lynch $\&$ Hunter in\cite{Palik1991}), magnesium sulfide (MnS;\cite{Huffman1967}), sodium sulfide (Na$_2$S;\cite{Morley2012}), zinc sulfide (ZnS;\cite{Querry1987}), and potassium chloride (KCL; Palick $\&$ Addamiano in\cite{Palik1998}) (ordered by condensation temperature). The third aerosol regime consists of condensed ices that form deep cloud decks in solar system planets and potentially cooler Y-dwarfs ($\leq$ 400K)\cite{Morley2014,Mang2022}. These ices could cause the opaque cloud deck found in our retrievals above, which are then lofted to higher atmospheric pressures to cause the observed scattering slope, or they could condense \emph{in situ} in the colder upper atmosphere: water ice (H$_2$O;\cite{Warren1984}), ammonia ice (NH$_3$;\cite{Martonchik1984}), and methane ice (CH$_4$;\cite{Martonchik1994}) (ordered by condensation temperature).

We find that all aerosol species, with the exception of MnS and hexene, provide good fits to the scattering slope and only imprint weak absorption features into the transmission spectrum. Via Bayesian model comparisons, the best-fit haze and soot species is the water-rich organic Exohaze (the 400K variant), the best-fit T-Y dwarf cloud species is KCl and the best-fit ice is NH$_3$. Of these three aerosols, KCl has the highest Bayesian evidence. The potential presence of KCl would be consistent with expectations for cold T-Y dwarf models\cite{Morley2012}, where KCl forms the highest, low-density cloud. However, we note that a simple gray cloud deck + haze model (as used in the Main text) is preferred over KCl by $\sim$ 2\,$\sigma$.  Therefore, the present data for WD~1856~b is not sufficiently precise to identify a clear preference for which specific aerosols are present in WD~1856~b's atmosphere.

\begin{figure*}
    \centering
    \includegraphics[width=\textwidth]{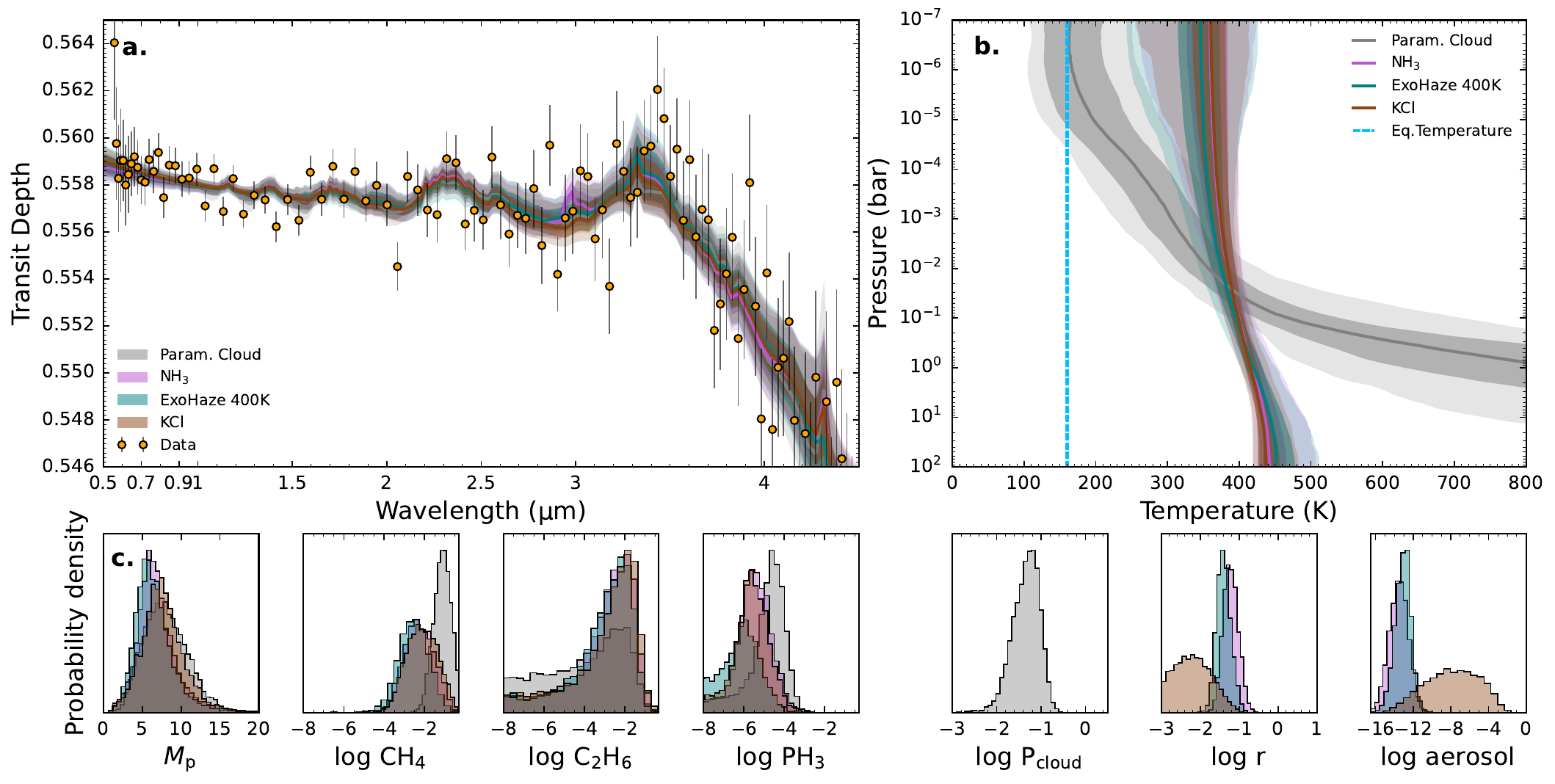}
    \caption{\textbf{Composition-specific Mie scattering retrievals}. \textbf{a.} Three retrieval models including Mie scattering (NH$_3$ ice, purple; ExoHaze at 400\,K, green; and KCl, orange) are compared to our reference non-Mie scattering grey cloud deck+haze model (grey). All four retrieval models produce similar fits to WD~1856~b's transmission spectrum (\firefly; orange circles with 1\,$\sigma$ error bars). \textbf{b.} Corresponding pressure-temperature profiles. The Mie scattering profiles are colder in the deeper atmosphere due to the inability of these models to produce an optically thick deep cloud deck. \textbf{c.} Corresponding posterior distributions for key atmospheric properties.}
    \label{fig:Mie_retirevals}
\end{figure*}

Our Mie scattering retrievals provide insights into the range of particle sizes and abundances compatible with WD~1856~b's short wavelength scattering slope (see Extended Data Figure~\ref{fig:Mie_retirevals}). The ExoHaze and NH$_3$ ice models favor a collection of small particles ($\sim$ 0.03\,$\mu$m) with low mixing ratios ($\sim 10^{-14}$), whereas the KCl model favors even smaller particles ($\sim$ 0.01 $\mu$m) with a higher abundance ($\sim 10^{-8}$). Compared with our default grey cloud deck + haze retrieval model, we find consistent results for other model parameters to within 1\,$\sigma$. In particular, we show that the retrieved planetary mass is not sensitive to the assumed aerosol model. We do find $\sim$ 1\,dex lower median CH$_4$ abundances for the Mie scattering retrievals, and hence a lower C/H ratio, but the CH$_4$ abundance distribution is still consistent with our results in the Main text. We note that the marginal evidence of C$_2$H$_6$ somewhat strengthens when including Mie scattering compared to the deck + haze model (see Extended Data Figure~\ref{fig:Mie_retirevals}), but this molecule is not significantly detected with the present data.

Our retrieved temperature structure from the Mie scattering retrievals also indicates an atmosphere significantly warmer than WD~1856~b's equilibrium temperature (Extended Data Figure~\ref{fig:Mie_retirevals}). As with our grey cloud and haze retrieval, we also find a temperature of $\sim$ 400\,K in the thermal photosphere near 10--100\,mbar. However, since the Mie scattering retrievals cannot produce an optically thick cloud deck at the pressures required to obscure thermal emission from the deep atmosphere ($\sim 10^{-1.5}$\,bar), the Mie retrievals compensate by making the P-T profile essentially isothermal in the deep atmosphere (i.e., the retrieved P-T profile shown in Figure~\ref{fig:Retrieved_spectrum_PT} in the Main text is more physical). We note that our uniform aerosol Mie scattering retrievals are incompatible with the P-T profile used in the main text\cite{Piette2020}, since a collection of small aerosols are not able to simultaneously block the deep adiabatic thermal flux and fit the scattering slope. The chosen P-T profile parameterisation here\cite{Madhusudhan2009} tends to favor a nearly isothermal upper atmosphere, which suffices for the exploration here of the aerosol properties consistent with WD~1856~b's scattering slope. Future explorations of WD~1856~b's cloud structure and radiative properties, such as composite cloud models with multiple scattering, are a rich area to deepen our understanding of WD~1856~b's atmosphere.
 
\subsection*{Evolution of the WD~1856 System}

\subsubsection*{Host Progenitor and White Dwarf}

We examined the evolution of the progenitor star of WD~1856 by consulting the MIST evolutionary models\cite{Choi2016} for non-rotating solar-metallicity stars in the appropriate mass range ($M_{\rm{progenitor}} = 1.36^{+0.29}_{-0.18} \, M_{\rm{Sun}}$). From these models, we extracted fiducial estimates of the main sequence lifetime ($4^{+2.4}_{-1.8}$\,Gyr) using an initial-final mass relation\cite{Cummings2018,Kiman2022}, the duration of the thermally pulsing AGB stage ($1.55^{+0.26}_{-0.10}$\,Myr) and the post-AGB/pre-white-dwarf stage ($0.034^{+0.053}_{-0.002}$\,Myr). The latter is here defined as the elapsed time between the final thermal pulse and the cooling of the exposed core to an effective temperature of 100,000\,K. This yields a total system age of $9.4^{+2.5}_{-1.9}$\,Gyr.

We calculated the cooling age of the white dwarf host by evolving MESA white dwarf models of the appropriate mass down to $T_{\rm eff}=4920\,$K. We used MESA r23.05.1\cite{Bauer2023}. This MESA release now includes carbon--oxygen fractionation\cite{Blouin2021}, which is important here as the white dwarf is in the process of crystallizing. A standard helium layer of $\log M_{\rm He}/M_{*}=-2$ was assumed, while a relatively thin hydrogen layer of $\log M_{\rm H}/M_{*}=-6$ was used. This is much thinner than the canonical value of $\log M_{\rm H}/M_{*}=-4$\cite{Renedo2010}, but is motivated by the fact that the model atmosphere analysis points to an atmosphere containing a mix of hydrogen and helium. This presumably requires the superficial convection zone to extend just below the hydrogen layer, thereby diluting hydrogen with helium. From this constraint we can estimate $\log M_{\rm H}/M_{*}=-6$\cite{Rolland2018}. Cooling calculations were performed for different carbon--oxygen core composition profiles to account for current model uncertainties: a standard profile predicted by stellar evolution\cite{Bauer2023} and an asteroseismologically derived stratification\cite{Giammichele2022} were used. We also calculated cooling models using different electron thermal conductivities\cite{Cassisi2007,Blouin2020} to account for current uncertainties at the transition between the regimes of moderate and strong degeneracy\cite{Cassisi2021}. From this analysis, we find a cooling age of $5.4 \pm 0.7\,$Gyr, where the uncertainty includes the systematic uncertainty sources listed above and a 2\% uncertainty on the star's $T_{\rm eff}$ and mass typical of white dwarfs in this temperature range\cite{Blouin2019}. This cooling age is consistent with estimates produced by other stellar evolution codes\cite{Renedo2010,Salaris2022}. 

\subsubsection*{Thermal History of WD~1856~b}

We reconstructed the thermal evolution of WD~1856~b under the assumption that the planet's cooling after migration has been similar to the cooling undergone by a substellar object after formation. We used cooling models from the ATMO2020\cite{Phillips2020} and Sonora Bobcat\cite{Marley2021} model grids. Each grid tabulates global quantities such as luminosity, effective temperature, radius, and surface gravity as a function of age for substellar objects of a given mass and bulk chemical composition, starting from an initial condition with high entropy. Both provide self-consistent evolutionary--atmospheric modeling frameworks, wherein the structure and evolution of the fully convective, adiabatic interior are computed with a cloudless, non-grey, rainout-chemical-equilibrium atmosphere as the surface boundary condition. The most important difference between ATMO2020 and Sonora Bobcat is that the former neglects some relevant opacity sources at effective temperatures above 2000 K, leading to faster cooling at high temperatures in the ATMO2020 models. ATMO2020 furnishes models of solar-metallicity objects, whereas Sonora Bobcat provides models for both solar-metallicity and somewhat metal-enriched (${\rm [M/H]}=+0.5$) objects. We consider these three sets of models in our analysis below.

Reconstructing the thermal history of WD~1856~b requires us to choose a model grid and specify three parameters: planetary mass ($M_{\rm p}$), current planetary effective temperature ($T_{\rm eff,p}$), and current white dwarf cooling age ($t_{\rm wd}$). For a given $M_{\rm p}$ and model grid, we obtained the effective temperature as a function of time by adding a uniform offset ($t_{0}$) to the model age ($t_{\rm p}$) such that the model temperature matches $T_{\rm eff,p}$ at $t_{\rm p} = t_{\rm wd} - t_{0}$ (using linear interpolation between the tabulated model ages and temperatures). The cooling models predict a high effective temperature ($\sim 1500$--$3000$\,K) at $t_{0}$; these values are plausible for planets that have been tidally heated during high-eccentricity migration\cite{Rozner2022, Glanz2022} or have survived a common-envelope phase\cite{Lagos2021, Merlov2021, O'Connor2023}. We therefore interpret $t_{0}$ as an estimate of the time of the planet's reheating during migration, expressed as a white dwarf cooling age. Because cooling is rapid at high $T_{\rm eff,p}$, our estimate of $t_{0}$ is robust to theoretical uncertainties in what temperature the planet should be immediately after migration.

We considered cooling models with $M_{\rm p}$ between 0.5 and 20\,$M_{\rm J}$, covering the range of samples from the atmospheric retrieval posterior distributions for the \firefly and \Juniper JWST data reductions. Each grid samples a finite number of mass values; when considering objects of arbitrary mass between grid points, we used the cooling model with the nearest mass on the grid. Both the ATMO2020 and Sonora Bobcat models are spaced by $\sim$0.5--1\,$M_{\rm J}$ in mass over the range we consider, so our approach does not introduce significant error in a given reconstruction compared to interpolating between adjacent models. 

We generated ensembles of possible thermal histories using the mass and effective temperature constraints derived from the NIRSpec PRISM transmission spectrum of WD~1856~b. Specifically, we considered nearly 10,000 values of $M_{\rm p}$ from the atmospheric retrieval posterior distribution alongside the nearly 10,000 corresponding values of $T_{\rm eff,p}$ obtained via the procedure described above for each data reduction (see `Atmospheric Retrieval Analysis'). Our samples of $M_{\rm p}$ and $T_{\rm eff,p}$ are not statistically independent, since each pair of values is derived from a single sample from the distribution of atmospheric models consistent with our NIRSpec PRISM transmission spectrum. This has significant bearing on the range of thermal histories we can infer from the data, because the cooling rate is a sensitive function of mass. 

\begin{figure*}
    \centering
    \includegraphics[width=\textwidth]{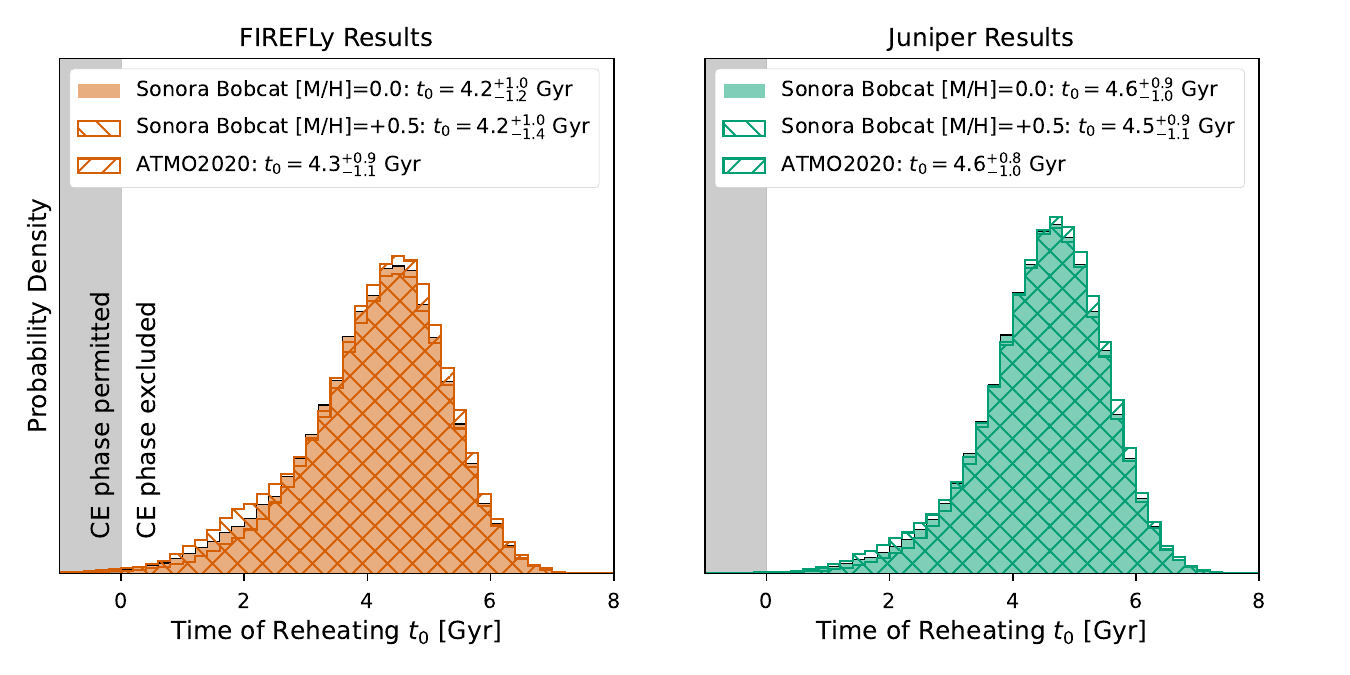}
    \caption{\textbf{The inferred time of WD~1856~b's reheating using different cooling models}. Each histogram contains $\approx$ 100,000 possible histories generated from the same set of properties ($(M_{\rm p}, T_{\rm eff,p}, t_{\rm wd})$, the former two from the atmospheric retrieval samples) but different cooling models (shading and hatching). The left panel corresponds to the \firefly reduction, the right is for \Juniper. The legend summarises the 1\,$\sigma$ credible region of $t_{0}$ for each cooling model. Only values of $t_{0} < 0$ (shaded regions) allow for a common envelope evolutionary phase.}
    \label{fig:t0_histograms}
\end{figure*}

On the other hand, our estimated $t_{\rm wd}$ is independent of our atmospheric retrieval analysis. For each pair of $M_{\rm p}$ and $T_{\rm eff,p}$ values, we generated 10 random values of $t_{\rm wd}$ drawn from a Gaussian distribution with mean $5.4$\,Gyr and standard deviation $0.7$\,Gyr. Each ensemble therefore comprises $\approx$ 100,000 possible thermal histories consistent with WD~1856~b's transmission spectrum. We generated one ensemble for each model grid. Extended Data Figure~\ref{fig:t0_histograms} shows the distribution of calculated $t_{0}$ values for the three cooling models and two data reductions, from which we derive a statistical constraint on $t_{0}$. The results reported in the Main text were obtained using the solar-metallicity Sonora Bobcat models (solid orange and green histograms; see also Figure~\ref{fig:thermal_evolution}). If we use the ATMO2020 models, we find a comparable $t_{0} = 4.3^{+0.9}_{-1.1}$\,Gyr for \firefly and $4.6^{+0.8}_{-1.0}$\,Gyr for \Juniper. Using the metal-enriched Sonora Bobcat models yields $t_{0} = 4.2^{+1.0}_{-1.4}$\,Gyr for \firefly and $4.5^{+0.9}_{-1.1}$\,Gyr for \Juniper. The conclusions we draw from modeling WD~1856~b's thermal evolution are therefore robust to both the JWST data reduction and the choice of cooling models, given the models currently available. 

In a small fraction of cases ($< 0.15\%$) for each ensemble, we calculate values $t_{0} < 0$. These correspond to the highest $M_{\rm p}$ values sampled from the atmospheric retrieval posterior. Negative values of $t_{0}$ arise in these cases because we have calculated $t_{0}$ by extrapolating the cooling models back to the effective temperatures expected among newborn brown dwarfs of $\sim 20 M_{\rm J}$ ($> 2000$\,K). However, these results are unphysical according to the interpretation of $t_{0}$ as the time elapsed between the end of the AGB phase and the planet's reheating/migration. If we stipulate that reconstructed thermal histories be truncated for $t_{0} < 0$, then these few cases are consistent with common-envelope evolution in that it is possible for the planet to have achieved its current temperature by passively cooling since the end of the AGB phase (albeit from a cooler, lower-entropy state than those implied in cases with $t_{0} > 0$). The fact remains, however, that the vast majority ($> 99.85\%$) of cases imply $t_{0}$ values that cannot coincide with a common-envelope phase in all three ensembles. Thus, we conclude that reheating during the white-dwarf phase (consistent with high-eccentricity migration) is preferred over reheating during common-envelope evolution at $> 2\,\sigma$ (for \firefly) and $> 3\,\sigma$ (for \Juniper). Further theoretical study is needed to corroborate or qualify this conclusion, as we describe below. 

\begin{figure*}
    \centering
    \includegraphics[width=\textwidth]{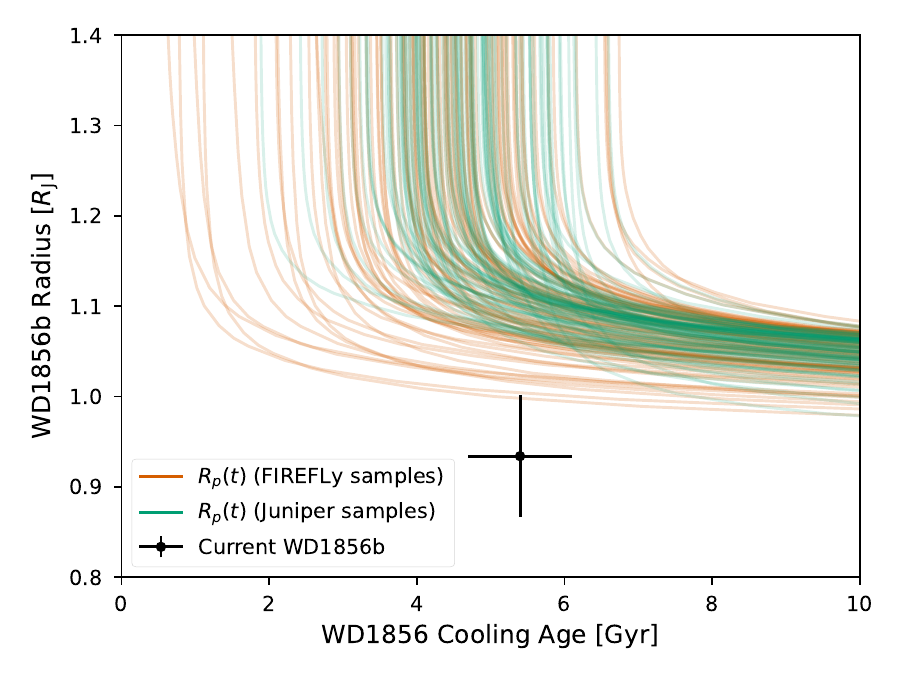}
    \caption{\textbf{Radius evolution of WD~1856~b.} The observed radius of WD~1856~b and white dwarf cooling age (black point and 1\,$\sigma$ error bars) and compared with the modelled radius  as a function of its age for \firefly (orange curves) and \Juniper (green curves). The models arise from the temperature-based reconstruction of WD~1856~b's thermal evolution.}
    \label{fig:WD1856b_radius_evolution}
\end{figure*}

Our method of reconstructing thermal histories is based on backward extrapolation of the effective temperature only. However, the cooling models also predict the evolution of WD~1856~b's radius; these predictions should agree in principle. In Extended Data Figure~\ref{fig:WD1856b_radius_evolution}, we show the radius evolution implied by our reconstruction method for 100 samples from the Sonora Bobcat ensemble for both data reductions. For the observed radius, we use the best-fitting value $R_{\rm p} = 0.911 \pm 0.020 R_{\rm J}$ from the \firefly reduction. We also include a systematic error of $\pm 0.050 R_{\rm J}$ given by the range of best-fitting radius values covered by the two data reductions, for a total uncertainty of $\pm 0.054 R_{\rm J}$. We see that many of the temperature-based reconstructions overestimate WD~1856~b's radius by $\approx$ 2\,$\sigma$. Future efforts to understand WD~1856~b's thermal evolution should reproduce both the effective temperature and radius. A clue as to the origin of this discrepancy comes from the heavy-element enrichment of WD~1856~b's envelope, suggested by our retrieved CH$_{4}$ abundance, as planetary radius decreases with increasing metallicity at a fixed mass and internal entropy. Model grids of comparable quality to ATMO2020 and Sonora Bobcat that are applicable to objects as massive ($\sim 7\,M_J$) and metal-rich ($\sim 100\,\times$\,solar) as WD~1856~b have not been developed or published to our knowledge.

We note that we neglected the irradiation of WD~1856~b by the host white dwarf in our reconstruction of the planet's thermal history. Irradiation is a key ingredient in modeling the structure and evolution of short-period exoplanets around main-sequence stars, such as hot Jupiters\cite{Guillot1996, Arras2006}. The importance of irradiation in the case of WD~1856~b can be gauged by calculating the ratio of the power emitted from the planet's photosphere to the power incident on the planet from the star: 
\begin{equation}
    \mathcal{R} = 4 \left( \frac{T_{\rm eff,p}}{T_{\rm eff,*}} \right)^{4} \left( \frac{a_{\rm orb}}{R_{*}} \right)^{2},
\end{equation}
where $T_{\rm eff,*}$ and $T_{\rm eff,p}$ are the host star and planet's measured effective temperatures, $R_{*}$ is the host radius, and $a_{\rm orb}$ is the orbital semi-major axis (assuming a near-circular orbit). Using the system parameters as determined in this work, we calculate $\mathcal{R} \approx 25$, indicating that the planet's self-luminosity overwhelms the power received from the star. Our reconstructed histories generally find that this ratio was larger in the past (except perhaps in the first several Myr after the white dwarf formed). Thus, we argue that irradiation has had a minor effect on WD~1856~b's previous thermal evolution. It would be of interest to self-consistently model the evolution of a substellar body with time-dependent irradiation, as would be the case in proximity to a cooling white dwarf. We leave this for future work. 

\subsubsection*{Alternatives to Reheating During Migration}

We considered several alternative explanations for the elevated effective temperature of WD~1856~b, all of which we deemed implausible or unlikely. We briefly describe each of them here along with our reasoning. 

First, WD~1856~b's observed effective temperature cannot be explained purely by passive cooling over the system's total age of $\sim$10\,Gyr\cite{Vanderburg2020}. This is readily ruled out by consulting theoretical cooling models\cite{Phillips2020, Marley2021}. In order to have an effective temperature of $\sim$400\,K at an age of 10\,Gyr, WD~1856~b would need to have a mass of $\sim 24 M_{\rm J}$. Our observations rule out such a high mass at $> 3 \sigma$ confidence.

WD~1856~b's mass may be above the threshold for deuterium fusion in its core ($\sim$13\,$M_{\rm J}$) within $2\sigma$. However, whilst it is possible that WD~1856~b was once heated internally by nuclear reactions, this cannot explain its present-day properties. Models of deuterium-burning brown dwarfs predict a total luminosity many orders of magnitude greater than that of WD~1856~b\cite{Chabrier2000}. The duration of deuterium burning ($\sim$3--50\,Myr depending on mass,\cite{Chabrier2000}) is drastically shorter than the system's total age, so WD~1856~b's primordial deuterium would have been destroyed early in the host's main-sequence lifetime. 

Due to WD~1856~b's proximity to its host, tidal interactions are another possible heat source inside the planet; this would be analogous to the heating of the Galilean satellite Io via its tidal interaction with Jupiter\cite{Peale1979}. For tidal heating to operate, WD~1856~b's orbit would need to be slightly eccentric, rather than circular as is typically assumed. Assuming that the power dissipated by tidal friction is equal to the total power emitted by WD~1856~b, we calculate the effective temperature of the planet, using the standard ``equilibrium tide'' theory\cite{Hut1981}, as: 
\begin{align}
    T_{\rm eff,p} &= \left( \frac{21}{8 \pi} \frac{G^{2} M_{*}^{3} R_{\rm p}^{3}}{\sigma_{\rm SB} a_{\rm orb}^{9}} k_{\rm 2p} \tau_{\rm p} e_{\rm orb}^{2} \right)^{1/4} \nonumber \\
    &\approx 400 \, {\rm K} \left( \frac{M_{*}}{0.60 M_{\odot}} \frac{R_{\rm p}}{0.91 R_{J}} \right)^{3/4} \left( \frac{a_{\rm orb}}{0.02 \, {\rm au}} \right)^{-9/4} \nonumber \\ 
    & \hspace{1cm} \times \left( \frac{k_{\rm 2p}}{0.4} \frac{\tau_{\rm p}}{0.1 \, {\rm s}} \right)^{1/4} \left( \frac{e_{\rm orb}}{0.02} \right)^{1/2}. \label{eq:Teff_tides}
\end{align}
Here, $G$ is the gravitational constant, $\sigma_{\rm SB}$ is the Stefan--Boltzmann constant, $M_{*}$ is the host white dwarf's mass, $R_{\rm p}$ is WD~1856~b's radius, and $a_{\rm orb}$ and $e_{\rm orb}$ are the orbital semi-major axis and eccentricity (with $e_{\rm orb} \ll 1$). The quantities $k_{\rm 2p}$ and $\tau_{\rm p}$, known respectively as the tidal Love number and tidal lag-time, characterize the dissipation inside WD~1856~b in the standard equilibrium tidal theory. The reference values of $M_{*}$, $R_{\rm p}$, and $a_{\rm orb}$ used in Eq.\ (\ref{eq:Teff_tides}) match the observed system parameters. For $k_{\rm 2p}$ and $\tau_{\rm p}$, we adopt values similar to those inferred for Jupiter's dissipation of the tide raised by Io\cite{Gavrilov1977, Lainey2009}. We see that tidal heating could, in principle, sustain WD~1856~b's observed effective temperature for an orbital eccentricity of $\approx 0.02$ (the highly uncertain values of $k_{\rm 2p}$ and $\tau_{\rm p}$ notwithstanding). This would be consistent with orbital circularization in the end-stage of high-eccentricity migration. However, the same dissipation would damp the orbital eccentricity on a characteristic timescale of $\sim$0.075~Gyr\cite{Hut1981}. In this picture, we are observing WD~1856~b in the very last, short-lived stage of high-eccentricity migration. Although we cannot rule it out based on the available data, we regard this explanation as unlikely. \\

\section*{Extended Data Tables and Figures}

\clearpage

\noindent \textbf{Data Availability}\\

\noindent The raw data from this study is available via the Space Science Telescope Institute's Mikulski Archive for Space Telescopes (\url{https://archive.stsci.edu/}) under program JWST-GO-2358. The \firefly and \Juniper transmission spectra of WD~1856~b are available on Zenodo at \url{https://doi.org/10.5281/zenodo.18200586}.
\\

\noindent \textbf{Code Availability}\\

\noindent \POSEIDON is available at \url{https://github.com/MartianColonist/POSEIDON}. A Python script to reproduce the retrieval results is available on Zenodo at \url{https://doi.org/10.5281/zenodo.18200586}.  \\
\noindent \texttt{batman} is available at \url{https://github.com/lkreidberg/batman}. \\
\noindent \texttt{emcee} is available at \url{https://github.com/dfm/emcee}. \\
\noindent \texttt{ExoTiC-LD} is available at \url{https://github.com/Exo-TiC/ExoTiC-LD}. \\
\noindent \texttt{MESA} is available at \url{https://zenodo.org/records/7983526}. \\
\noindent \firefly and \Juniper are not presently publicly available. Requests for additional details on \firefly and \Juniper should be addressed to \href{David K. Sing}{mailto:dsing@jhu.edu} and \href{Victoria A. Boehm}{mailto:vab55@cornell.edu}, respectively.

\vskip 1 cm
\backmatter

\noindent \textbf{Acknowledgments}\\

\noindent This version of the article has been accepted for publication, after peer review but is not the Version of Record and does not reflect post-acceptance improvements, or any corrections. The Version of Record is available online at: \textbf{http://dx.doi.org/[insert DOI]}.
This work is based on observations made with the NASA/ESA/CSA James Webb Space Telescope. The data were obtained from the Mikulski Archive for Space Telescopes at the Space Telescope Science Institute, which is operated by the Association of Universities for Research in Astronomy, Inc., under NASA contract NAS 5-03127 for JWST. Support for Program \#2358 was provided through a grant from STScI under NASA contract NAS5-03127.
R.J.M. acknowledges support from NASA through the NASA Hubble Fellowship grant HST-HF2-51513.001, awarded by the Space Telescope Science Institute, which is operated by the Association of Universities for Research in Astronomy, Inc., for NASA, under contract NAS 5-26555.
C.E.O. acknowledges support by the National Science Foundation under Grant No.\ AST-2107796 (PI: Dong Lai). 
L.A.P.~acknowledges research support from the NSF Graduate Research Fellowship.  This material is based upon work supported by the National Science Foundation Graduate Research Fellowship Program under Grant No. DGE-1746060 and the NSF INTERN Program under Grant No. DGE-2137419.
T.O.F. acknowledges support from NASA through the NASA FINESST grant 80NSSC22K1893.
S.B. acknowledges support from the Canadian Institute for Theoretical Astrophysics. 
The authors thank the anonymous referees for their valuable feedback that improved the quality  of this study.
\\

\noindent \textbf{Author Contributions}\\

\noindent R.J.M. led the overall team efforts, designed the JWST GO-2358 Program, conducted the retrieval analysis, coordinated the paper writing and Figure creation, and contributed to the text. 
C.E.O. led the thermal evolution and migration history interpretation and contributed to the text.
V.A.B. led the development of \Juniper, its application to the NIRSpec PRISM data reduction, and contributed to the text.
E.M.M. contributed to the data reduction analysis and to the text.
D.S. led the \firefly reduction analysis and contributed to the text.
E.M. conducted the Mie scattering retrieval analysis and contributed to the text.
L.C.M and L.A.P contributed to the atmospheric analysis and contributed to the text.
T.O.F contributed to the development of \Juniper, and contributed to the text.
S.B. performed the white dwarf host spectral analysis and the white dwarf cooling calculations.
N.K.L., J.V, N.E.B, M.L., J.D.L, M.S.M, I.M., and S.E.M. contributed to the writing of the paper.
All co-authors read and agreed with the conclusions of the manuscript.\\

\noindent \textbf{Competing Interests} The authors declare no competing interests.\\

\noindent\textbf{Additional information}\newline
\textbf{Correspondence and requests for materials} should be addressed to \href{Ryan J. MacDonald}{mailto:ryan.macdonald@st-andrews.ac.uk}.\newline
\textbf{Reprints and permissions information} is available at \url{www.nature.com/reprints}.


\clearpage

\begin{figure*}
    \centering
    \includegraphics[width=\textwidth]{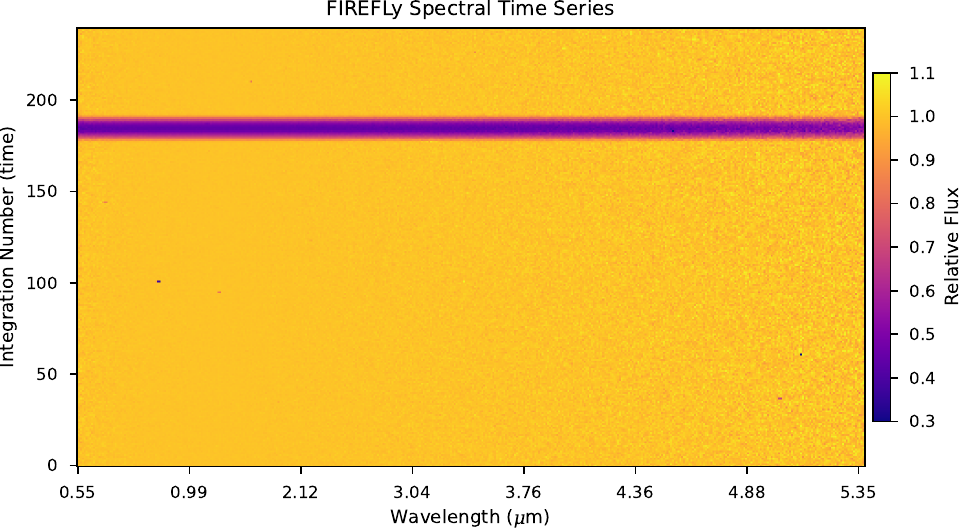}
    \caption{\textbf{NIRSpec PRISM light curves for WD~1856~b's transit.} The spectrophotometric transit light curves from the \firefly data reduction show the relative flux of the WD~1856 system as a function of wavelength and time.}
    \label{fig:2dlightcurvemap}
\end{figure*}

\clearpage

\begin{table*}
    \centering
    \caption{\textbf{Best-fitting transit model parameters.} The white light curve transit model parameters for the \Juniper and \firefly data reductions are compared. We note that the planet-star radius ratio ($R_p/R_*$) and the impact parameter ($b$) are strongly correlated for a grazing transit. The period is fixed for both reductions and taken from Ref.\cite{Kubiak2023}.}
    \begin{tabular}{|c|c|c|}
        \hline
        Parameter & \Juniper & \firefly \\ \hline
        $a/R_*$ & 339.25 +/- 5.92 & 330.72 +/- 3.50 \\ \hline
        $R_p/R_*$ & 7.54 +/- 0.28 & 7.03 +/- 0.16 \\ \hline
        $b$ & 7.34 +/- 0.20 & 7.19 +/- 0.09 \\ \hline
        Period {(}days, fixed{)} & 1.407939217 \cite{Kubiak2023} & 1.407939217 \cite{Kubiak2023} \\ \hline
    \end{tabular}
    \label{tab:data_fit_params}
\end{table*}

\begin{table*}
    \centering
    \caption{\textbf{Atmospheric retrieval model parameters and priors.} The free parameters defining the \POSEIDON retrieval model used in the Main text are listed. The volume mixing ratio parameters encode the chemical abundances of CH$_4$, NH$_3$, H$_2$O, CO$_2$, CO, HCN, C$_2$H$_2$, C$_2$H$_4$, C$_2$H$_6$, H$_2$S, and PH$_3$. The prior for each parameter is uniformly distributed between a lower and upper limit, $\mathcal{U}(x_{1}, x_{2})$.}
    \begin{tabular}{|llc|}
        \hline
        Parameter & Description & Priors \\
        \hline \\[-12pt]
        \textbf{Planetary Properties} & & \\
        \hspace{1pt} ${\rm R_{p, ref}}$ ($R_{J}$) & Reference planet radius & $\mathcal{U}(0.8451, 1.0329)$ \\
        \hspace{1pt} ${\rm M_{p}}$ ($M_{J}$) & Planetary mass & $\mathcal{U}(0.5, 20.0)$ \\[2pt]
        \textbf{P-T Profile} & & \\
        \hspace{1pt} $T_{\rm{phot}}$ (K) & Temperature at 0.1 bar & $\mathcal{U}(100, 2000)$ \\
        \hspace{1pt} $\Delta T_{1}$ (K) & Temperature change: $10^{-6}$--$10^{-5}$\,bar & $\mathcal{U}(0, 200)$ \\
        \hspace{1pt} $\Delta T_{2}$ (K) & Temperature change: $10^{-5}$--$10^{-4}$\,bar & $\mathcal{U}(0, 200)$ \\
        \hspace{1pt} $\Delta T_{3}$ (K) & Temperature change: $10^{-4}$--$10^{-3}$\,bar & $\mathcal{U}(0, 200)$ \\
        \hspace{1pt} $\Delta T_{4}$ (K) & Temperature change: $10^{-3}$--$10^{-2}$\,bar & $\mathcal{U}(0, 500)$ \\
        \hspace{1pt} $\Delta T_{5}$ (K) & Temperature change: $10^{-2}$--$10^{-1}$\,bar & $\mathcal{U}(0, 500)$ \\
        \hspace{1pt} $\Delta T_{6}$ (K) & Temperature change: $10^{-1}$--1\,bar & $\mathcal{U}(0, 1000)$ \\
        \hspace{1pt} $\Delta T_{7}$ (K) & Temperature change: 1--$10^{1}$\,bar & $\mathcal{U}(0, 1500)$ \\
        \hspace{1pt} $\Delta T_{8}$ (K) & Temperature change: $10^{1}$--$10^{2}$\,bar & $\mathcal{U}(0, 2000)$ \\[2pt]
        \textbf{Composition} & & \\
        \hspace{1pt} $\log_{10} X_{i}$ & Volume mixing ratio of $i^{\rm{th}}$ gas & $\mathcal{U}(-12, -0.3)$ \\[2pt]
        \textbf{Aerosols} & & \\
        \hspace{1pt} $\log_{10} a$ & Rayleigh enhancement factor & $\mathcal{U}(-4, 8)$ \\
        \hspace{1pt} $\gamma$ & Scattering slope & $\mathcal{U}(-20, 2)$ \\
        \hspace{1pt} $\log_{10} (P_{\rm cloud} / \rm{bar})$ & Cloud top pressure & $\mathcal{U}(-6, 2)$ \\
        \hline
    \end{tabular}
    \label{tab:retrieval_priors}
\end{table*}

\begin{table*}
    \centering
    \caption{\textbf{Atmospheric retrieval model statistics.} Model statistics are listed for the 25-parameter reference \POSEIDON model (Ref.), alongside nested models without nightside thermal emission, without aerosols, without hydrocarbons, and without CH$_4$ for both the \firefly and \Juniper data reductions. The model statistics are the number of free parameters ($N_{\rm{param}}$), the log-Bayesian evidence ($\ln \mathcal{Z}$), the log-Bayes factor ($\ln \mathcal{B}$), the equivalent detection significance (Det. Sig.), the chi-squared of the best-fit spectrum ($\chi^2_{\rm{min}}$), the degrees of freedom (d.o.f.), and the reduced chi-squared of the best-fit spectrum ($\chi^2_{\mathrm{\nu, \, min}}$). }
    \begin{tabular}{|lccccccc|}
        \hline
        Data + Model & $N_{\rm{param}}$ & $\ln \mathcal{Z}$ & $\ln \mathcal{B}$ & Det. Sig. & $\chi^2_{\rm{min}}$ & d.o.f. & $\chi^2_{\nu, \, \rm{min}}$ \\
        \hline \\[-12pt]
        \textbf{\firefly} & & & & & & & \\
        \hspace{1pt} Ref. & 25 & 595.2 & Ref. & Ref. & 105.3 & 102 & 1.03 \\
        \hspace{1pt} No Emission & 25 & 449.3 & 145.9 & 17.3$\sigma$ & 348.5 & 102 & 3.42 \\
        \hspace{1pt} No Aerosols & 22 & 582.9 & 12.3 & 5.3$\sigma$ & 125.8 & 105 & 1.20 \\
        \hspace{1pt} No Hydrocarbons & 21 & 590.0 & 5.12 & 3.6$\sigma$ & 125.1 & 106 & 1.18 \\
        \hspace{1pt} No CH$_4$ & 24 & 592.3 & 2.87 & 2.9$\sigma$ & 111.1 & 103 & 1.08 \\[2pt]
        \textbf{\Juniper} & & & & & & & \\
        \hspace{1pt} Ref. & 25 & 577.16 & Ref. & Ref. & 130.5 & 102 & 1.28 \\
        \hspace{1pt} No Emission & 25 & 408.2 & 169.0 & 18.5$\sigma$ & 415.6 & 102 & 4.07 \\
        \hspace{1pt} No Aerosols & 22 & 562.7 & 14.5 & 5.7$\sigma$ & 155.1 & 105 & 1.48 \\
        \hspace{1pt} No Hydrocarbons & 21 & 568.6 & 8.59 & 4.5$\sigma$ & 151.1 & 106 & 1.43 \\
        \hspace{1pt} No CH$_4$ & 24 & 573.8 & 3.40 & 3.1$\sigma$ & 133.7 & 103 & 1.30 \\
        \hline
    \end{tabular}
    \label{tab:retrieval_model_stats}
\end{table*}

\end{document}